\documentclass[12pt]{article}
\usepackage{amssymb,amsmath,epsfig}
\allowdisplaybreaks

\begin{document}

\title{\bf Existence of Static Wormholes in $f(\mathcal{G},T)$ Gravity}
\author{M. Sharif \thanks {msharif.math@pu.edu.pk} and Ayesha Ikram
\thanks{ayeshamaths91@gmail.com}\\
Department of Mathematics, University of the Punjab,\\
Quaid-e-Azam Campus, Lahore-54590, Pakistan.}

\date{}
\maketitle

\begin{abstract}
This paper investigates static spherically symmetric traversable
wormhole solutions in $f(\mathcal{G},T)$ gravity ($\mathcal{G}$ and
$T$ represent the Gauss-Bonnet invariant and trace of the
energy-momentum tensor, respectively). We construct explicit
expressions for ordinary matter by taking specific form of red-shift
function and $f(\mathcal{G},T)$ model. To analyze possible existence
of wormholes, we consider anisotropic, isotropic as well as
barotropic matter distributions. The graphical analysis shows the
violation of null energy condition for the effective energy-momentum
tensor throughout the evolution while ordinary matter meets energy
constraints in certain regions for each case of matter distribution.
It is concluded that traversable WH solutions are physically
acceptable in this theory.
\end{abstract}

\section{Introduction}

Gauss-Bonnet (GB) invariant has a significant importance in higher
dimensional theories as well as in describing the early and
late-times cosmic evolution. It is a quadratic curvature invariant
of the form
$\mathcal{G}=R_{\alpha\beta\gamma\delta}R^{\alpha\beta\gamma\delta}
-4R_{\alpha\beta}R^{\alpha\beta}+R^{2}$, where
$R_{\alpha\beta\gamma\delta},~R_{\alpha\beta}$ and $R$ represent the
Riemann tensor, Ricci tensor and Ricci scalar, respectively. This
quadratic invariant is a four-dimensional topological term and is
free from spin-2 ghost instabilities \cite{1}. Nojiri and Odintsov
\cite{2} added the generic function $f(\mathcal{G})$ in the
Einstein-Hilbert action (dubbed as $f(\mathcal{G})$ gravity) to
explore the dynamics of GB invariant in four dimensions. This
modified theory is consistent with solar system constraints as well
as endowed with a quite rich cosmological structure \cite{3}. It is
interesting to investigate the effects of non-minimal coupling
between curvature and matter on cosmic evolution. Recently, we have
established this curvature-matter coupling in the action of
$f(\mathcal{G})$ gravity named as $f(\mathcal{G},T)$ gravity
\cite{4}. We have found that energy-momentum tensor is not conserved
due to the presence of non-minimally curvature-matter coupling. This
non-conservation produces an extra force due to which dust particles
move along geodesic lines of geometry while non-geodesic
trajectories are followed by massive particles. The background of
cosmological evolutionary models corresponding to
phantom/non-phantom eras, power-law solutions as well as de Sitter
universe can be discussed in this theory \cite{5}.

A wormhole (WH) is defined as a hypothetical bridge or tunnel that
provides a shortcut across the spacetime for long distances.
Inter-universe WH allows a path of communication between distant
patches of distinct spacetimes while a subway connecting distant
regions of the same spacetime is dubbed as an intra-universe WH. The
simplest solution of the Einstein field equations representing this
hypothetical connecting shortcut is Schwarzschild WH also known as
Einstein-Rosen bridge \cite{6}. Schwarzschild WH does not allow
two-way travel (non-traversable WH) due to the presence of strong
tidal gravitational forces at WH throat which would destroy anything
that tries to pass through. Moreover, it evolves with time such that
expansion (circumference increases from zero to finite) and
contraction (shrinks to zero) of WH throat are very rapid and it
does not allow anything to pass through the tunnel. Schwarzschild WH
possesses highly unstable antihorizon that changes to a horizon even
when light passes through it, thereby closing the WH throat. To
overcome these drawbacks, Morris and Thorne \cite{7} gave the
concept of traversable WHs. They observed that these WHs must be
sustained by the matter which violates the null energy condition
(NEC) dubbed as exotic matter. The presence of this unrealistic form
of matter pushes the walls of WH apart and prevents the WH throat to
shrink.

The search for alternative source of violation such that ordinary
matter meets the energy conditions has always been a subject of
great interest. Brane WHs, dynamical WH solutions, non-commutative
geometry, generalized Chaplygin gas, modified theories of gravity,
etc provide a source that helps to minimize the usage of exotic
matter to support the WH geometry \cite{8}. In modified theories of
gravity, the effective energy-momentum tensor (includes
higher-curvature terms and ordinary matter variables) acts as a
source of violation required for the traversability and thus
provides a possibility for the existence of realistic WH. Furey and
DeBenedicts \cite{9} investigated WH solutions for $R^{2}$ and
$R^{-1}$ theories of gravity and found the positivity of weak energy
condition (WEC) in the neighborhood of WH throat. Lobo and Oliveira
\cite{10} discussed static spherically symmetric WH solutions and
found that realistic WH geometries threaded by ordinary matter can
be formed in $f(R)$ gravity.

Azizi \cite{11} explored WH geometries and found that NEC is
satisfied for barotropic matter configuration in $f(R,T)$ gravity.
The physically acceptable WH solutions are observed for barotropic
fluid in the background of $f(R)$ gravity \cite{11a}. Sharif and
Rani \cite{12} studied static spherically symmetric WH solutions for
exponential as well as logarithmic forms of generalized teleparallel
gravity and found that WEC is violated in galactic halo region for
both models. Mehdizadeh et al. \cite{13} discussed realistic
traversable WH solutions in Einstein GB gravity. We have explored WH
geometries for traceless, isotropic as well as barotropic matter
distributions and concluded that realistic WH solutions exist in
$f(\mathcal{G})$ gravity only for radial barotropic fluid \cite{14}.
Zubair and his collaborators \cite{15} investigated WH solutions and
found that stable physically acceptable WH solutions exist for
anisotropic matter configuration.

In this paper, we explore static spherically symmetric WH solutions
for anisotropic, isotropic and barobropic matter distributions in
$f(\mathcal{G},T)$ gravity. The paper has the following format. In
the next section, we discuss basic concepts related to
$f(\mathcal{G},T)$ gravity, WH geometry as well as energy
conditions. Section \textbf{3} is devoted to construct WH solutions
using three types of fluid for specific $f(\mathcal{G},T)$ model. In
section \textbf{4}, we summarize the results.

\section{Field Equations for Wormhole Construction}

The action for $f(\mathcal{G},T)$ gravity is defined as \cite{4}
\begin{equation}\label{1a}
\mathcal{I}=\int\left(\frac{R+f(\mathcal{G},T)}{2\kappa^2}
+\mathcal{L}_{m}\right)\sqrt{-g}d^{4}x,
\end{equation}
where $\mathcal{L}_{m},~\kappa^2$ and $g$ denote the Lagrangian
density associated with matter configuration, coupling constant and
determinant of the metric tensor, respectively. Varying the action
(\ref{1a}) with respect to $g_{\alpha\beta}$, we obtain the fourth
order field equations as follows
\begin{equation}\label{2a}
G_{\alpha\beta}=\kappa^2T_{\alpha\beta}^{\mathrm{eff}}=\kappa^2(T_{\alpha\beta}
+T_{\alpha\beta}^{\mathcal{G}T}),
\end{equation}
where $G_{\alpha\beta}$ and $T_{\alpha\beta}^{\mathrm{eff}}$
represent Einstein tensor and effective energy-momentum tensor,
respectively. The expression for $T_{\alpha\beta}^{\mathcal{G}T}$ is
given by
\begin{eqnarray}\nonumber
\kappa^2T_{\alpha\beta}^{\mathcal{G}T}&=&\frac{1}{2}
g_{\alpha\beta}f(\mathcal{G},T)-T_{\alpha\beta}f_{T}(\mathcal{G},T)
-\Theta_{\alpha\beta}f_{T}(\mathcal{G},T)-2RR_{\alpha\beta}f_{\mathcal{G}}
(\mathcal{G},T)\\\nonumber&+&4R_{\alpha}^{~\zeta}R_{\zeta\beta}f_{\mathcal{G}}
(\mathcal{G},T)+4R_{\alpha\gamma\beta\delta}R^{\gamma\delta}
f_{\mathcal{G}}(\mathcal{G},T)-2R_{\alpha}^{~\gamma\delta\zeta}
R_{\beta\gamma\delta\zeta}f_{\mathcal{G}}(\mathcal{G},T)
\\\nonumber&-&2Rg_{\alpha\beta}\Box f_{\mathcal{G}}
(\mathcal{G},T)-4R_{\alpha}^{~\zeta}\nabla_{\beta}\nabla_{\zeta}
f_{\mathcal{G}}(\mathcal{G},T)-4R_{\beta}^{~\zeta}\nabla_{\alpha}
\nabla_{\zeta}f_{\mathcal{G}}(\mathcal{G},T)\\\nonumber&+&2R
\nabla_{\alpha}\nabla_{\beta}f_{\mathcal{G}}(\mathcal{G},T)+4g_{\alpha\beta}
R^{\gamma\delta}\nabla_{\gamma}\nabla_{\delta}f_{\mathcal{G}}(\mathcal{G},T)
+4R_{\alpha\beta}\Box f_{\mathcal{G}}(\mathcal{G},T)
\\\nonumber&-&4R_{\alpha\gamma\beta\delta}\nabla^{\gamma}
\nabla^{\delta}f_{\mathcal{G}}(\mathcal{G},T),
\end{eqnarray}
where $f_{T}(\mathcal{G},T)=\partial f(\mathcal{G},T)/\partial
T,~f_{\mathcal{G}}(\mathcal{G},T)=\partial f(\mathcal{G},T)/\partial
\mathcal{G},~\Box=\nabla^{2}=\nabla_{\alpha}\nabla^{\alpha}$ and
$\nabla_{\alpha}$ is a covariant derivative whereas tensor
$\Theta_{\alpha\beta}$ has the form \cite{16}
\begin{equation}\nonumber
\Theta_{\alpha\beta}=g^{\gamma\zeta}\frac{\delta
T_{\gamma\zeta}}{\delta
g_{\alpha\beta}}=g_{\alpha\beta}\mathcal{L}_{m}-2g^{\gamma\zeta}
\frac{\partial^{2}\mathcal{L}_{m}}{\partial g^{\alpha\beta}\partial
g^{\gamma\zeta}}-2T_{\alpha\beta},
\end{equation}
with the assumption that the matter Lagrangian density depends only
on $g_{\alpha\beta}$. The energy-momentum tensor for anisotropic
matter distribution is
\begin{equation}\label{3a}
T_{\alpha\beta}=(\rho+P_{t})u_{\alpha}u_{\beta}-P_{t}g_{\alpha\beta}
+(P_{r}-P_{t})\eta_{\alpha}\eta_{\beta},
\end{equation}
where $u_{\alpha},~\eta_{\alpha},~\rho,~P_{r}$ and $P_{t}$ represent
the four velocity, unit four-vector in radial direction, energy
density, radial and tangential pressures of the fluid, respectively.
For anisotropic configuration, we take $\mathcal{L}_{m}=\rho$, the
resultant expression for $\Theta_{\alpha\beta}$ becomes
\begin{equation}\label{4a}
\Theta_{\alpha\beta}=\rho g_{\alpha\beta}-2T_{\alpha\beta}.
\end{equation}

The static spherically symmetric line element describing the
geometry of traversable WH is given by \cite{7}
\begin{equation}\label{5a}
ds^2=e^{2\psi(r)}dt^2-\left(1-\frac{\chi(r)}{r}\right)^{-1}dr^2-r^2
d\theta^2-r^2\sin^2\theta d\phi^2,
\end{equation}
where $\psi(r)$ and $\chi(r)$ are the generic functions of $r$ known
as red-shift and shape functions, respectively. The first function
$\psi(r)$ measures the magnitude of gravitational red-shift of a
photon while geometry of WH is determined by $\chi(r)$. For the
traversability of WH, $\psi(r)$ must be finite everywhere to satisfy
the no-horizon condition. The shape function must represent the
increasing behavior with respect to $r$ such that $(1-\chi(r)/r)>0$
throughout the tunnel to maintain the WH geometry. In addition, the
value of $\chi(r)$ and $r$ must be same at throat, i.e.,
$\chi(r_{\mathrm{th}})=r_{\mathrm{th}}$. The fundamental condition
known as flaring-out condition, given by
$(\psi(r)-\psi'(r)r)/\psi^2(r)>0$, needs to be satisfied throughout
the evolution while the constraint $\chi'(r_{\mathrm{th}})<1$ should
be imposed at the throat. Also, the asymptotically flatness
condition, $\chi(r)/r\rightarrow0$ as $r\rightarrow\infty$, must be
fulfilled.

Using Eqs.(\ref{3a})-(\ref{5a}) in (\ref{2a}), we obtain the
following set of field equations
\begin{eqnarray}\nonumber
\kappa^2\rho&=&\frac{\chi'}{r^2}-\frac{1}{2}f(\mathcal{G},T)
+\frac{1}{2}\mathcal{G}f_{\mathcal{G}}(\mathcal{G},T)+\frac{2}{r^4}
(r\chi'-\chi)\left(2-\frac{3\chi}{r}\right)f_{\mathcal{G}}'
(\mathcal{G},T)\\\label{6a}&+&\frac{4\chi}{r^3}\left(1-\frac{\chi}{r}\right)
f_{\mathcal{G}}''(\mathcal{G},T),\\\nonumber\kappa^2P_{r}&=&
\left[\frac{2\psi'}{r}\left(1-\frac{\chi}{r}\right)-\frac{\chi}{r^3}
+\frac{1}{2}f(\mathcal{G},T)-\frac{1}{2}\mathcal{G}f_{\mathcal{G}}
(\mathcal{G},T)-\frac{4\psi'}{r^3}(3\chi-2r)\right.\\\label{7a}&\times&\left.
\left(1-\frac{\chi}{r}\right)f_{\mathcal{G}}'(\mathcal{G},T)-\rho
f_{T}(\mathcal{G},T)\right]\left(1+\frac{f_{T}(\mathcal{G},T)}{\kappa^2}
\right)^{-1},\\\nonumber\kappa^2P_{t}&=&\left[\left(\frac{\psi'}
{r}+\psi'^{2}+\psi''\right)\left(1-\frac{\chi}{r}\right)+\frac{1}{2r^{3}}
(1+r\psi')(\chi-\chi'r)+\frac{1}{2}f(\mathcal{G},T)\right.
\\\nonumber&-&\left.\frac{1}{2}\mathcal{G}f_{\mathcal{G}}(\mathcal{G},T)
+\frac{2}{r}\left(1-\frac{\chi}{r}\right)\left\{2(\psi'^{2}+\psi'')
\left(1-\frac{\chi}{r}\right)-\frac{3\psi'}{r^2}(\chi'r\right.\right.
\\\nonumber&-&\left.\left.\chi)\right\}f_{\mathcal{G}}'(\mathcal{G},T)
+\frac{4\psi'}{r}\left(1-\frac{\chi}{r}\right)^2f_{\mathcal{G}}''
(\mathcal{G},T)-\rho f_{T}(\mathcal{G},T)\right]\\\label{8a}&\times&
\left(1+\frac{f_{T}(\mathcal{G},T)}{\kappa^2}\right)^{-1},
\end{eqnarray}
where prime is the derivative with respect to $r$ and
$T=\rho-P_{r}-2P_{t}$ whereas GB invariant has the following
expression
\begin{equation}\label{9a}
\mathcal{G}=\frac{4}{r^4}\left[\psi'(\chi-\chi'r)\left(2
-\frac{3\chi}{r}\right)-2r\chi(\psi'^{2}+\psi'')
\left(1-\frac{\chi}{r}\right)\right].
\end{equation}
The above system of equations (\ref{6a})-(\ref{8a}) shows that the
generic function $f(\mathcal{G},T)$ has a direct dependence on
matter variables therefore, it would be difficult to find the
explicit expressions for $\rho,~P_{r}$ and $P_{t}$. The favorable
approach to solve this system for matter contents is to choose
$f(\mathcal{G},T)=F(\mathcal{G})+\mathcal{F}(T)$ with
$\mathcal{F}(T)=\Upsilon T$, where $\Upsilon$ is an arbitrary
constant. This simplest choice of $f(\mathcal{G},T)$ function does
not involve the direct curvature-matter non-minimally coupling and
is considered as the correction to $f(\mathcal{G})$ gravity. For
this particular $f(\mathcal{G},T)$ form, we simplify the equations
(\ref{6a})-(\ref{8a}) as follows
\begin{eqnarray}\label{10a}
\rho&=&\frac{1}{2(1+2\Upsilon)}\left[\left(\frac{2+5\Upsilon}
{1+\Upsilon}\right)\Omega_{1}+\Upsilon(\Omega_{2}+2\Omega_{3})
\right],\\\label{11a}P_{r}&=&\frac{-1}{2(1+2\Upsilon)}\left[
\left(\frac{\Upsilon}{1+\Upsilon}\right)\Omega_{1}-(2+3\Upsilon)
\Omega_{2}+2\Upsilon\Omega_{3}\right],\\\label{12a}P_{t}&=&\frac{-1}{2
(1+\Upsilon)(1+2\Upsilon)}\left[\Upsilon\Omega_{1}+\Upsilon(1-\Upsilon)
\Omega_{2}+2(1-\Upsilon)^2\Omega_{3}\right],
\end{eqnarray}
where $\kappa^{2}=1$ and
\begin{eqnarray}\nonumber
\Omega_{1}&=&\frac{\chi'}{r^2}-\frac{1}{2}F(\mathcal{G})
+\frac{1}{2}\mathcal{G}F_{\mathcal{G}}(\mathcal{G})+\frac{2}{r^4}
(r\chi'-\chi)\left(2-\frac{3\chi}{r}\right)F_{\mathcal{G}}'
(\mathcal{G})\\\nonumber&+&\frac{4\chi}{r^3}\left(1-\frac{\chi}{r}\right)
F_{\mathcal{G}}''(\mathcal{G}),\\\nonumber\Omega_{2}&=&
\frac{1}{(1+\Upsilon)}\left[\frac{2\psi'}{r}\left(1-\frac{\chi}{r}\right)
-\frac{\chi}{r^3}+\frac{1}{2}F(\mathcal{G})-\frac{1}{2}\mathcal{G}
F_{\mathcal{G}}(\mathcal{G})-\frac{4\psi'}{r^3}(3\chi-2r)\right.
\\\nonumber&\times&\left.\left(1-\frac{\chi}{r}\right)
F_{\mathcal{G}}'(\mathcal{G})\right],\\\nonumber\Omega_{3}&=&
\frac{1}{(1+\Upsilon)}\left[\left(\frac{\psi'}{r}+\psi'^{2}
+\psi''\right)\left(1-\frac{\chi}{r}\right)+\frac{1}{2r^3}(1+r\psi')
(\chi-\chi'r)\right.\\\nonumber&+&\left.\frac{1}{2}F(\mathcal{G})
-\frac{1}{2}\mathcal{G}F_{\mathcal{G}}(\mathcal{G})+\frac{2}{r}\left(1
-\frac{\chi}{r}\right)\left\{2(\psi'^{2}+\psi'')\left(1-\frac{\chi}{r}
\right)-\frac{3\psi'}{r^2}\right.\right.\\\nonumber&\times&\left.\left.
(\chi'r-\chi)\right\}F_{\mathcal{G}}'(\mathcal{G})+\frac{4\psi'}{r}
\left(1-\frac{\chi}{r}\right)^2F
_{\mathcal{G}}''(\mathcal{G})\right].
\end{eqnarray}
It is worth mentioning here that the above equations reduce to
$f(\mathcal{G})$ gravity for $\Upsilon=0$ while general relativity
(GR) is recovered when the contribution of generic function
vanishes, i.e., $F(\mathcal{G})=0$ with $\Upsilon=0$ \cite{14}.

Energy conditions are used to discuss physically realistic matter
configuration that are originated from Raychaudhuri equations. These
constraints are imposed on the energy-momentum tensor and possess an
interesting feature that they are coordinate invariant. The
Raychaudhuri equations describe the temporal evolution of expansion
scalar $(\theta)$ for the congruences of timelike $(v^{\alpha})$ and
null $(l_{\alpha}$) geodesics as \cite{17}
\begin{eqnarray}\nonumber
&&\frac{d\theta}{d\tau}-\omega_{\alpha\beta}\omega^{\alpha\beta}
+\sigma_{\alpha\beta}\sigma^{\alpha\beta}+\frac{1}{3}\theta^2
+R_{\alpha\beta}v^{\alpha}v^{\beta}=0,\\\nonumber&&\frac{d\theta}
{d\tau}-\omega_{\alpha\beta}\omega^{\alpha\beta}+\sigma_{\alpha
\beta}\sigma^{\alpha\beta}+\frac{1}{2}\theta^2+R_{\alpha\beta}
l^{\alpha}l^{\beta}=0,
\end{eqnarray}
where $\omega_{\alpha\beta}$ and $\sigma_{\alpha\beta}$ represent
the rotation and shear tensors, respectively. For non-geodesic
(timelike or null) congruences, the temporal evolution for expansion
scalar changes in the presence of acceleration term as \cite{18}
\begin{equation}\label{13a}
\frac{d\theta}{d\tau}-\omega_{\alpha\beta}\omega^{\alpha\beta}
+\sigma_{\alpha\beta}\sigma^{\alpha\beta}+\frac{1}{3}\theta^2
+R_{\alpha\beta}v^{\alpha}v^{\beta}-\mathcal{A}=0,
\end{equation}
the auxiliary term
$\mathcal{A}=\nabla_{\alpha}(u^{\beta}\nabla_{\beta}u^{\alpha})$
represents divergence of four-acceleration dubbed as acceleration
term which appears due to the non-gravitational force (pressure
gradient). Using the condition of attractive nature of gravity
($\theta<0$) and neglecting the quadratic terms, the Raychaudhuri
equations for non-geodesic congruences reduce to
\begin{equation}\nonumber
R_{\alpha\beta}v^{\alpha}v^{\beta}-\mathcal{A}\geq0,\quad
R_{\alpha\beta}l^{\alpha}l^{\beta}-\mathcal{A}\geq0.
\end{equation}
In terms of energy-momentum tensor, the above inequalities take the
form
\begin{equation}\label{14a}
\left(T_{\alpha\beta}-\frac{1}{2}g_{\alpha\beta}T\right)
v^{\alpha}v^{\beta}-\mathcal{A}\geq0,\quad\left(T_{\alpha\beta}
-\frac{1}{2}g_{\alpha\beta}T\right)l^{\alpha}l^{\beta}-\mathcal{A}\geq0.
\end{equation}

In modified theories of gravity, these inequalities are obtained by
replacing $T_{\alpha\beta}$ with $T_{\alpha\beta}^{\mathrm{eff}}$
since Raychaudhuri equations possess purely geometric nature. In
$f(\mathcal{G},T)$ gravity, we find that massive test particles
follow the non-geodesic trajectories due to the presence of an extra
force \cite{4}. The above inequalities (\ref{14a}) with
$T_{\alpha\beta}^{\mathrm{eff}}$ provide the null, weak, strong
(SEC) and dominant (DEC) energy conditions as
\begin{itemize}
\item NEC:$\quad\rho^{\mathrm{eff}}+P_{i}^{\mathrm{eff}}-\mathcal{A}
\geq0,~i=1,2,3$,\item WEC:$\quad\rho^{\mathrm{eff}}
+P_{i}^{\mathrm{eff}}-\mathcal{A}\geq0,~\rho^{\mathrm{eff}}
-\mathcal{A}\geq0,$\item SEC:$\quad\rho^{\mathrm{eff}}
+P_{i}^{\mathrm{eff}}-\mathcal{A}\geq0,~\rho^{\mathrm{eff}}
+\sum_{i}P_{i}^{\mathrm{eff}}-\mathcal{A}\geq0,$\item
DEC:$\quad\rho^{\mathrm{eff}}\pm P_{i}^{\mathrm{eff}}
-\mathcal{A}\geq0,~\rho^{\mathrm{eff}}-\mathcal{A}\geq0,$
\end{itemize}
where $\rho^{\mathrm{eff}}$ and $P^{\mathrm{eff}}$ represent the
effective energy density and pressure, respectively. The null energy
condition is considered as the fundamental energy bound whose
violation leads to the violation of all energy constraints. It is
worth mentioning here that the non-geodesic energy bounds in GR can
be obtained by replacing $\rho^{\mathrm{eff}}$ and
$P^{\mathrm{eff}}$ with usual matter contents $\rho$ and $P$,
respectively. In the absence of acceleration term, i.e., for
geodesic congruences, one can recover the usual energy conditions in
$f(\mathcal{G},T)$ gravity \cite{4}.

For the traversability of WH, the basic property is the violation of
NEC in GR. This violation prevents the WH throat to shrink and leads
to the physically unrealistic WH solutions. The modified theories of
gravity provide $T_{\alpha\beta}^{\mathrm{eff}}$ as an alternative
source to meet the violation of NEC. In this regard, these theories
may have an opportunity for usual matter configuration to fulfil the
energy constraints. Using the field equations (\ref{2a}), we obtain
NEC in $f(\mathcal{G},T)$ gravity
\begin{equation}\label{15a}
\rho^{\mathrm{eff}}+P_{r}^{\mathrm{eff}}-\mathcal{A}=\left(\frac{r\chi'-\chi}
{r^3}\right)\left(1-\frac{r\psi'}{2}\right)-\left(\psi'^{2}+\psi''\right)
\left(1-\frac{\chi}{r}\right),
\end{equation}
where the acceleration term for Eq.(\ref{5a}) is given by
\begin{equation}\label{16a}
\mathcal{A}=\left(1-\frac{\chi}{r}\right)\left[\psi''+\psi'^{2}
+\frac{2\psi'}{r}\right]-\frac{\psi'}{2r^2}\left(r\chi'-\chi\right).
\end{equation}
In the absence of acceleration term, the expression (\ref{15a})
becomes identical to $f(\mathcal{G})$ gravity \cite{14}.

\section{Wormhole Solutions}

In this section, we investigate WH solutions by taking three
different types of matter distribution for specific form of
$\psi(r)$ and viable $F(\mathcal{G})$ model. We assume the finite
red-shift function as \cite{19}
\begin{equation}\label{1b}
\psi(r)=-\frac{\lambda}{r}, \quad\lambda>0,
\end{equation}
which meets the no-horizon condition as well as shows asymptotically
flatness behavior at large distances, i.e., $\psi(r)\rightarrow0$ as
$r\rightarrow\infty$. The expression for GB invariant (\ref{9a})
takes the form
\begin{equation}\label{2b}
\mathcal{G}=\frac{4\lambda}{r^7}\left[(2r-3\chi)(\chi-r\chi')
-2\chi(\lambda-2r)\left(1-\frac{\chi}{r}\right)\right].
\end{equation}
We consider the following algebraic power-law form as \cite{20}
\begin{equation}\label{3b}
F(\mathcal{G})=\mu\mathcal{G}^{a}(1+\nu\mathcal{G}^{b}),
\end{equation}
where $a,~b,~\mu$ and $\nu$ are arbitrary constants. Under certain
conditions on these model parameters, this realistic model does not
possess any four types of finite-time future singularities as well
as efficiently describes the current cosmic acceleration. Using
Eqs.(\ref{1b}) and (\ref{3b}), the expressions for $\Omega_{i}$'s in
the field equations (\ref{10a})-(\ref{12a}) become
\begin{eqnarray}\nonumber
\Omega_{1}&=&\frac{\chi'}{r^2}-\frac{1}{2}\mu\mathcal{G}^{a}
(1+\nu\mathcal{G}^{b})+\frac{1}{2}\mu\mathcal{G}^{a}[a+\mu(a+b)
\mathcal{G}^{b}]+\frac{2\mu}{r^4}(r\chi'-\chi)\\\nonumber&\times&
\left(2-\frac{3\chi}{r}\right)\left[a(a-1)+\nu(a+b)(a+b-1)
\mathcal{G}^{b}\right]\mathcal{G}^{a-2}\mathcal{G}'
+\frac{4\chi}{r^3}\left(1-\frac{\chi}{r}\right)\\\nonumber&\times&
\left[a\mu(a-1)\mathcal{G}^{a-2}\left\{\mathcal{G}''+(a-2)
\frac{\mathcal{G}'^{2}}{\mathcal{G}}\right\}+\mu\nu(a+b)(a+b-1)
\mathcal{G}^{a+b-2}\right.\\\label{4b}&\times&\left.\left\{
\mathcal{G}''+(a+b-2)\frac{\mathcal{G}'^{2}}{\mathcal{G}}
\right\}\right],\\\nonumber\Omega_{2}&=&\frac{1}{(1+\Upsilon)}
\left[\frac{2\lambda}{r^3}\left(1-\frac{\chi}{r}\right)
-\frac{\chi}{r^3}+\frac{1}{2}\mu\mathcal{G}^{a}(1+\nu
\mathcal{G}^{b})-\frac{1}{2}\mu\mathcal{G}^{a}[a+\nu(a+b)
\right.\\\nonumber&\times&\left.\mathcal{G}^{b}]-\frac{4\mu\lambda}
{r^5}(3\chi-2r)\left(1-\frac{\chi}{r}\right)\left\{a(a-1)
+\nu(a+b)(a+b-1)\mathcal{G}^{b}\right\}\right.\\\label{5b}&\times&\left.
\mathcal{G}^{a-2}\mathcal{G}'\right],\\\nonumber
\Omega_{3}&=&\frac{1}{(1+\Upsilon)}\left[\frac{\lambda}{r^4}(\lambda-r)
\left(1-\frac{\chi}{r}\right)+\frac{1}{2r^4}(r+\lambda)(\chi-r\chi')
+\frac{1}{2}\mu\mathcal{G}^{a}(1+\nu\mathcal{G}^{b})\right.
\\\nonumber&-&\left.\frac{1}{2}\mu\mathcal{G}^{a}[a+\nu(a+b)
\mathcal{G}^{b}]+\frac{2\lambda\mu}{r^5}\left(1-\frac{\chi}{r}
\right)\left\{2(\lambda-2r)\left(1-\frac{\chi}{r}\right)
\right.\right.\\\nonumber&-&\left.\left.3(r\chi'-\chi)\right\}
\left[a(a-1)+\nu(a+b)(a+b-1)\mathcal{G}^{b}\right]
\mathcal{G}^{a-2}\mathcal{G}'\right.\\\nonumber&+&\left.
\frac{4\lambda}{r^3}\left(1-\frac{\chi}{r}\right)^{2}
\left\{a\mu(a-1)\mathcal{G}^{a-2}\left(\mathcal{G}''+(a-2)
\frac{\mathcal{G}'^{2}}{\mathcal{G}}\right)+\mu\nu(a+b)\right.\right.
\\\label{6b}&\times&\left.\left.(a+b-1)\mathcal{G}^{a+b-2}
\left(\mathcal{G}''+(a+b-2)\frac{\mathcal{G}'^{2}}{\mathcal{G}}
\right)\right\}\right].
\end{eqnarray}
In the following subsections, we analyze possible WH solutions for
anisotropic, isotropic as well as barotropic matter distributions.

\subsection{Anisotropic Fluid}

We consider the specific form of shape function as
\begin{eqnarray}\label{1c}
\chi(r)=r_{\mathrm{th}}\left(\frac{r_{\mathrm{th}}}{r}\right)^{n},
\end{eqnarray}
where $n$ is an arbitrary constant. This satisfies all the necessary
conditions of shape function for the existence of traversable WH.
The asymptotically flatness condition is fulfilled for this specific
form of $\chi(r)$. At the throat, the condition
$\chi(r_{\mathrm{th}})=r_{\mathrm{th}}$ trivially holds while the
validity of $\chi'(r_{\mathrm{th}})<1$ is achieved for $n>-1$. Lobo
and Oliveira \cite{10} investigated WH solutions in $f(R)$ gravity
for $n=-1/2,~1$ with this form of $\chi(r)$. Zubair and his
collaborators \cite{15} studied static spherically symmetric WH
solutions for $n=1/2,~1$ and $-3$ in the background of $f(R,T)$
gravity. For $n=1/2$, Pavlovic and Sossich \cite{21} explored the
existence of WH geometries for different forms of generic $f(R)$
function. Using Eq.(\ref{1c}) in (\ref{4b})-(\ref{6b}), we have
\begin{eqnarray}\nonumber
\Omega_{1}&=&-\frac{n}{r^2}\left(\frac{r_{\mathrm{th}}}{r}\right)^{n+1}
-\frac{1}{2}\mu\mathcal{G}^{a}(1+\nu\mathcal{G}^{b})+\frac{1}{2}\mu
\mathcal{G}^{a}[a+\nu(a+b)\mathcal{G}^{b}]\\\nonumber&-&\frac{2\mu}{r^3}
(n+1)\left(\frac{r_{\mathrm{th}}}{r}\right)^{n+2}\left[2-3\left(\frac
{r_{\mathrm{th}}}{r}\right)^{n+1}\right]\left[a(a-1)+\nu(a+b)\right.
\\\nonumber&\times&\left.(a+b-1)\mathcal{G}^{b}\right]\mathcal{G}^{a-2}
\mathcal{G}'+\frac{4}{r^2}\left(\frac{r_{\mathrm{th}}}{r}\right)^{n+1}
\left[1-\left(\frac{r_{\mathrm{th}}}{r}\right)^{n+1}\right]\left[a\mu
(a-1)\right.\\\nonumber&\times&\left.\mathcal{G}^{a-2}\left\{\mathcal{G}''
+(a-2)\frac{\mathcal{G}'^{2}}{\mathcal{G}}\right\}+\mu\nu(a+b)(a+b-1)
\mathcal{G}^{a+b-2}\right.\\\label{2c}&\times&\left.\left\{\mathcal{G}''
+(a+b-2)\frac{\mathcal{G}'^{2}}{\mathcal{G}}\right\}\right],\\\nonumber
\Omega_{2}&=&\frac{1}{(1+\Upsilon)}\left[\frac{2\lambda}{r^3}\left\{1-
\left(\frac{r_{\mathrm{th}}}{r}\right)^{n+1}\right\}-\frac{1}{r^2}\left(
\frac{r_{\mathrm{th}}}{r}\right)^{n+1}+\frac{1}{2}\mu\mathcal{G}^{a}(1+\nu
\mathcal{G}^{b})\right.\\\nonumber&-&\left.\frac{1}{2}\mu\mathcal{G}^{a}
[a+\nu(a+b)\mathcal{G}^{b}]-\frac{4\mu\lambda}{r^4}\left\{3\left(
\frac{r_{\mathrm{th}}}{r}\right)^{n+1}-2\right\}\left\{1-\left(
\frac{r_{\mathrm{th}}}{r}\right)^{n+1}\right\}\right.\\\label{3c}&\times&
\left.\left\{a(a-1)+\nu(a+b)(a+b-1)\mathcal{G}^{b}\right\}\mathcal{G}^{a-2}
\mathcal{G}'\right],\\\nonumber\Omega_{3}&=&\frac{1}{(1+\Upsilon)}
\left[\frac{\lambda}{r^4}(\lambda-r)\left\{1-\left(\frac{r_{\mathrm{th}}}{r}
\right)^{n+1}\right\}+\frac{1}{2r^3}(r+\lambda)(n+1)\left(\frac{r_{\mathrm{th}}}
{r}\right)^{n+1}\right.\\\nonumber&+&\left.\frac{1}{2}\mu\mathcal{G}^{a}
(1+\nu\mathcal{G}^{b})-\frac{1}{2}\mu\mathcal{G}^{a}[a+\nu(a+b)\mathcal{G}^{b}]
+\frac{2\mu\lambda}{r^5}\left\{1-\left(\frac{r_{\mathrm{th}}}{r}\right)^{n+1}
\right\}\right.\\\nonumber&\times&\left.\left[2(\lambda-2r)\left\{1-\left(
\frac{r_{\mathrm{th}}}{r}\right)^{n+1}\right\}+3r_{\mathrm{th}}(n+1)\left(
\frac{r_{\mathrm{th}}}{r}\right)^{n}\right][a(a-1)\right.\\\nonumber&+&\left.
\nu(a+b)(a+b-1)\mathcal{G}^{b}]\mathcal{G}^{a-2}\mathcal{G}'+\frac{4\lambda}
{r^3}\left\{1-\left(\frac{r_{\mathrm{th}}}{r}\right)^{n+1}\right\}^{2}\left[a
\mu(a-1)\right.\right.\\\nonumber&\times&\left.\left.\mathcal{G}^{a-2}\left\{
\mathcal{G}''+(a-2)\frac{\mathcal{G}'^{2}}{\mathcal{G}}\right\}+\mu\nu(a+b)
(a+b-1)\mathcal{G}^{a+b-2}\right.\right.\\\label{4c}&\times&\left.\left.
\left\{\mathcal{G}''+(a+b-2)\frac{\mathcal{G}'^{2}}{\mathcal{G}}\right\}
\right]\right],
\end{eqnarray}
where the expression for GB invariant (\ref{2b}) is given by
\begin{eqnarray}\nonumber
\mathcal{G}=\frac{4\lambda}{r^6}\left(\frac{r_{\mathrm{th}}}{r}
\right)^{n+1}\left[r(n+1)\left\{2-3\left(\frac{r_{\mathrm{th}}}{r}
\right)^{n+1}\right\}-2(\lambda-2r)\left\{1-\left(\frac{r_{\mathrm{th}}}
{r}\right)^{n+1}\right\}\right].
\end{eqnarray}
Substituting Eq.(\ref{1c}) in (\ref{15a}), the non-geodesic NEC in
$f(\mathcal{G},T)$ gravity reduces to
\begin{equation}\label{5c}
\rho^{\mathrm{eff}}+P_{r}^{\mathrm{eff}}-\mathcal{A}=\frac{\lambda}{r^3}
\left(2-\frac{\lambda}{r}\right)\left[1-\left(\frac{r_{\mathrm{th}}}
{r}\right)^{n+1}\right]-\frac{(n+1)}{r^2}\left(1-\frac{\lambda}{r}\right)
\left(\frac{r_{\mathrm{th}}}{r}\right)^{n+1}.
\end{equation}

The violation of this energy bound for different values of $n$ is
shown in Figure \textbf{1} with $\lambda=0.01$ and
$r_{\mathrm{th}}=1$. This violation provides a possibility to search
for physically acceptable WH solutions in the presence of
anisotropic matter distribution. For this purpose, we take
appropriate values of parameters to check the behavior of energy
conditions for ordinary matter. Figure \textbf{2} shows the validity
of NEC as well as WEC throughout the evolution for all three
considered values of $n$ with $\mu=1,~\nu=-1,~a=0.2$ and $b=0.25$.
This choice of model parameters corresponds to the case
$a>0,~b>0,~a\neq1$ and $b\neq1$ restricted with $0<a+b<1/2$ and
$\mu\nu<0$ to avoid all four types of finite-time future
singularities \cite{20}. The behavior of both energy conditions for
the cases $a>0,~b<0$ and $a\neq1$ with $0<a<1/2$ and $\mu<0$ is
given in Figure \textbf{3}. We set $\mu=-1,~\nu=1,~a=0.2$ and
$b=-0.25$ as an example and find that these model parameters meet
the energy bounds in the region $2\leq r\leq 10$.
\begin{figure}\centering
\epsfig{file=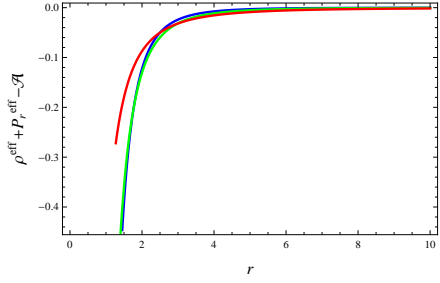, width=0.42\linewidth}\caption{Plots of NEC
versus $r$ for $n=1$ (blue), $0.5$ (red) and $-0.5$ (green).}
\end{figure}
\begin{figure}\centering
\epsfig{file=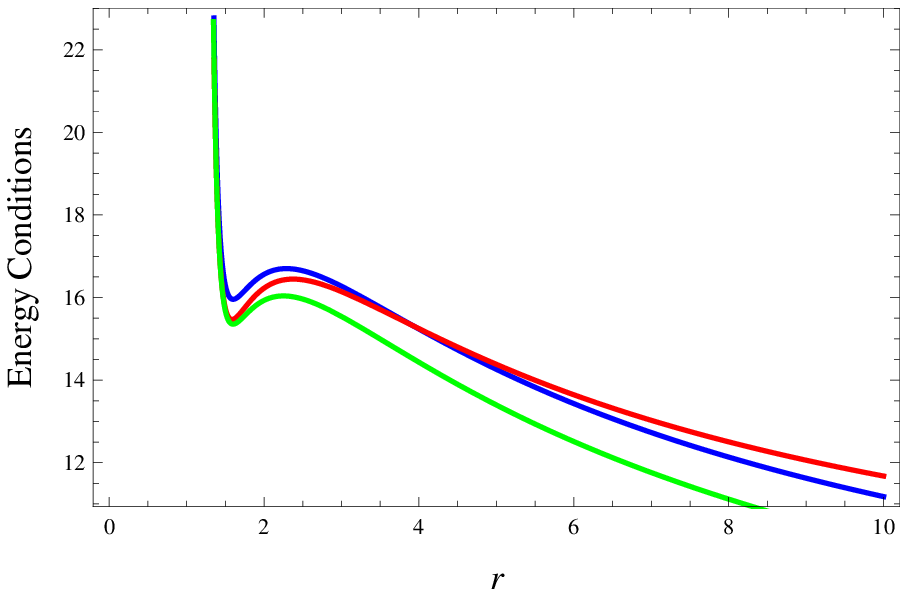, width=0.42\linewidth}\epsfig{file=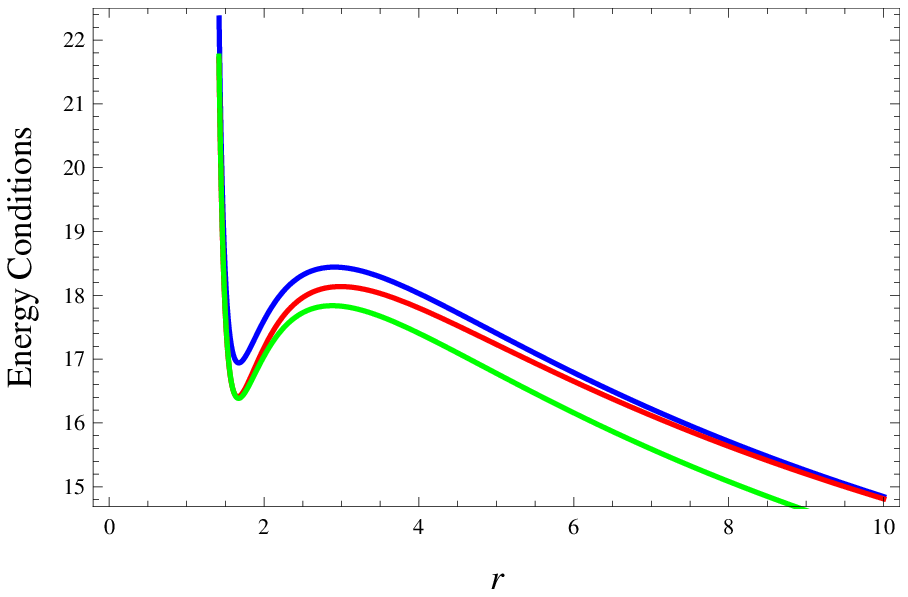,
width=0.42\linewidth}\\\epsfig{file=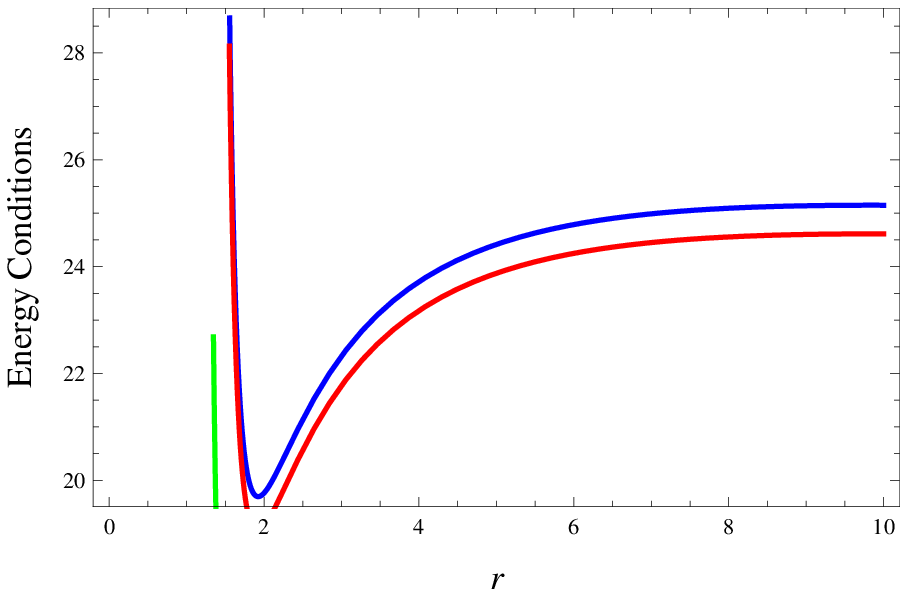,
width=0.42\linewidth}\caption{Plots of $\rho-\mathcal{A}$ (blue),
$\rho+P_{r}-\mathcal{A}$ (red) and $\rho+P_{t}-\mathcal{A}$ (green)
versus $r$ with $\Upsilon=0.05,~\mu=1,~\nu=-1,~a=0.2$ and $b=0.25$.
The upper case for $n=1$ (left) and $n=0.5$ (right) while lower case
for $n=-0.5$.}
\end{figure}
\begin{figure}\centering
\epsfig{file=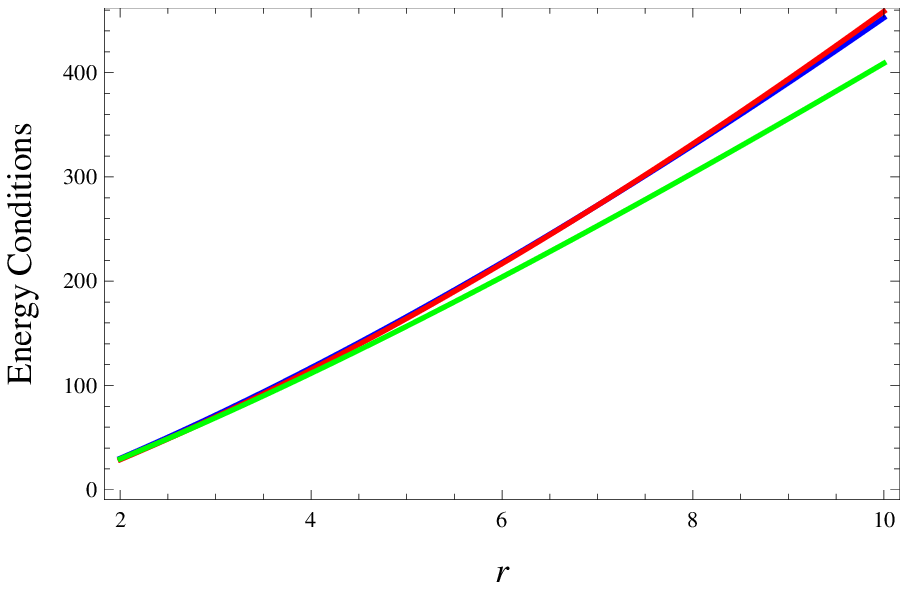, width=0.42\linewidth}\epsfig{file=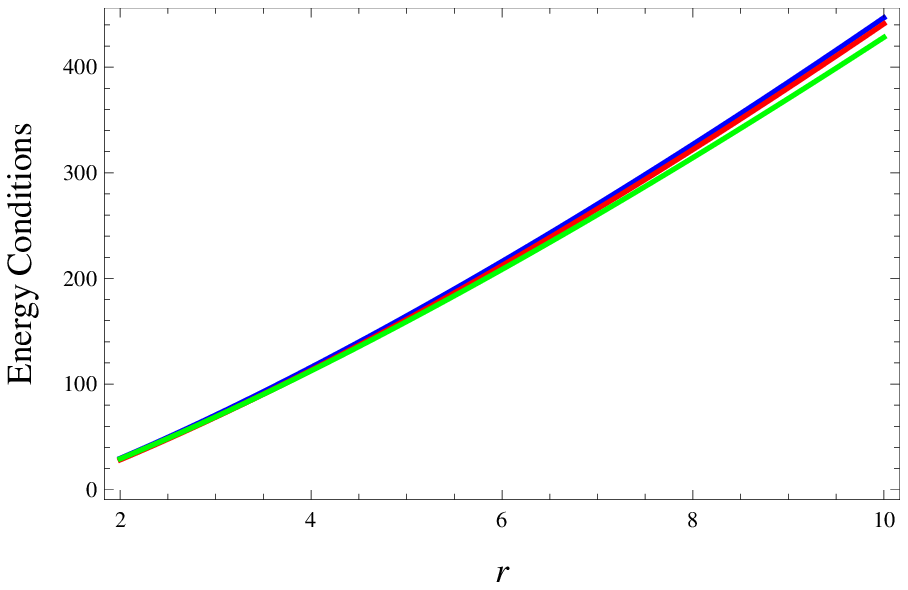,
width=0.42\linewidth}\\\epsfig{file=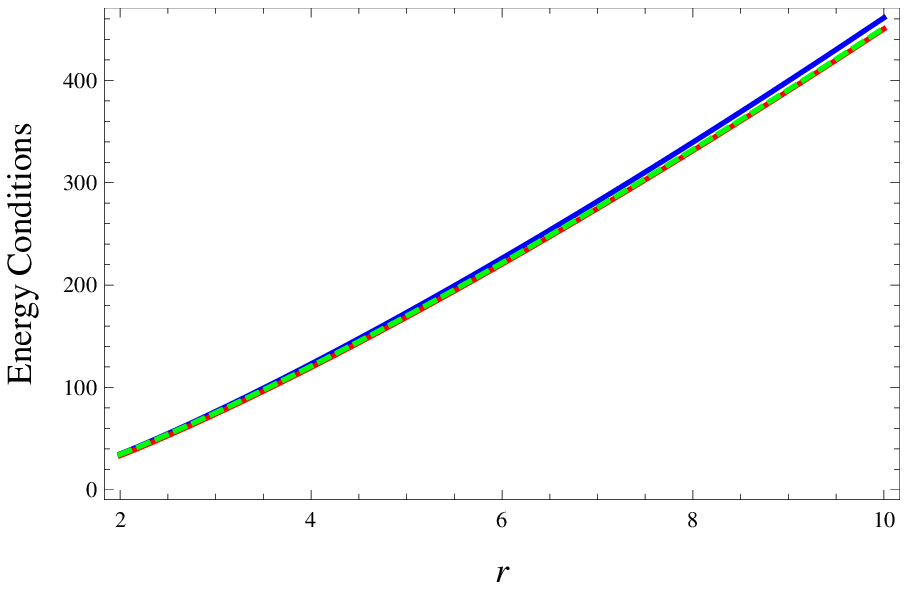,
width=0.42\linewidth}\caption{Plots of $\rho-\mathcal{A}$ (blue),
$\rho+P_{r}-\mathcal{A}$ (red) and $\rho+P_{t}-\mathcal{A}$ (green)
versus $r$ with $\Upsilon=0.05,~\mu=-1,~\nu=1,~a=0.2$ and $b=-0.25$.
The upper case for $n=1$ (left) and $n=0.5$ (right) while lower case
for $n=-0.5$.}
\end{figure}

The third possibility to avoid the finite-time future singularities
is generally described by $a<0,~b>0$ and $b\neq1$ while the
constraint $a+b<1/2$ with $\mu\nu<0$ is also imposed on model
parameters. In this case, we arbitrarily choose the values
$\mu=-1,~\nu=1,~a=-0.01$ and $b=0.25$ to analyze the validity of
energy conditions given in Figure \textbf{4}. It is observed that
the realistic WH solutions exist in the regions $3.2\leq r\leq
10,~3.5\leq r\leq 10$ and $4\leq r\leq 10$ for $n=1,~0.5$ and
$-0.5$, respectively. Thus, the physically acceptable region
decreases as the value of $n$ decreases. Figure \textbf{5} shows
that NEC as well as WEC remain positive in the region $4\leq r\leq
10$ with the choice of model parameters $a<0,~b<0$ and $\mu<0$
particularly for $\mu=-1,~\nu=1,~a=-0.01$ and $b=-0.25$. In this
case, WH geometries are also sustained by normal matter.
\begin{figure}\centering
\epsfig{file=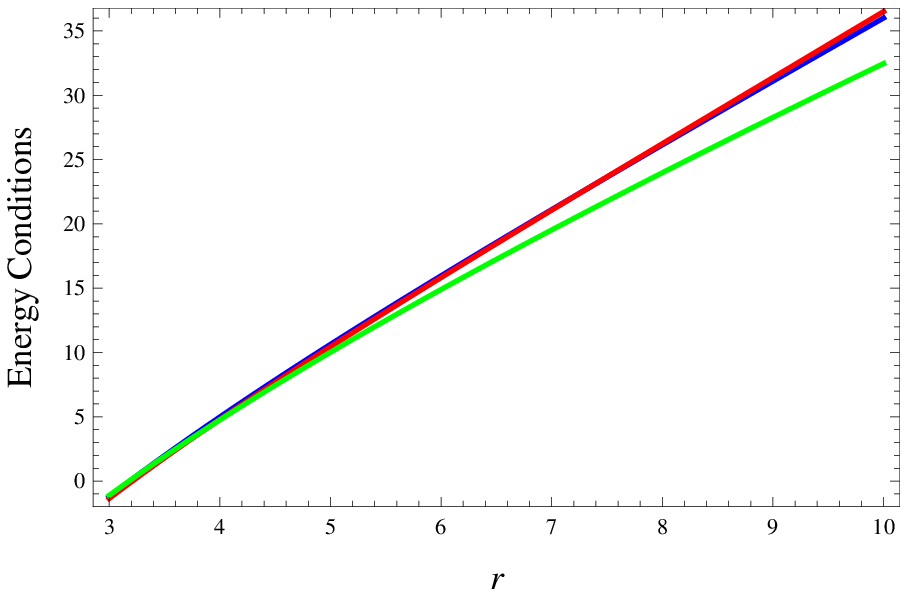, width=0.42\linewidth}\epsfig{file=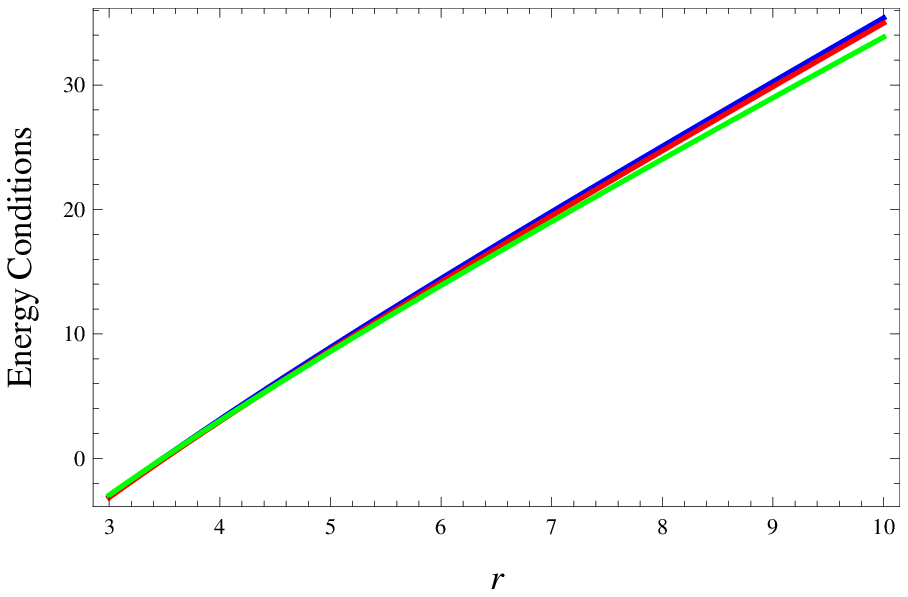,
width=0.42\linewidth}\\\epsfig{file=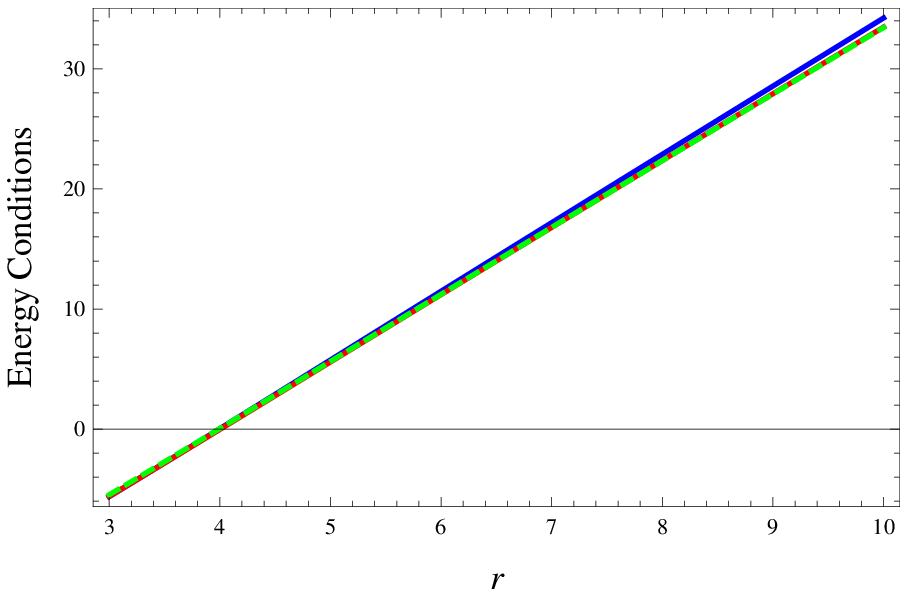,
width=0.42\linewidth}\caption{Plots of $\rho-\mathcal{A}$ (blue),
$\rho+P_{r}-\mathcal{A}$ (red) and $\rho+P_{t}-\mathcal{A}$ (green)
versus $r$ with $\Upsilon=0.05,~\mu=-1,~\nu=1,~a=-0.01$ and
$b=0.25$. The upper case for $n=1$ (left) and $n=0.5$ (right) while
lower case for $n=-0.5$.}
\end{figure}
\begin{figure}\centering
\epsfig{file=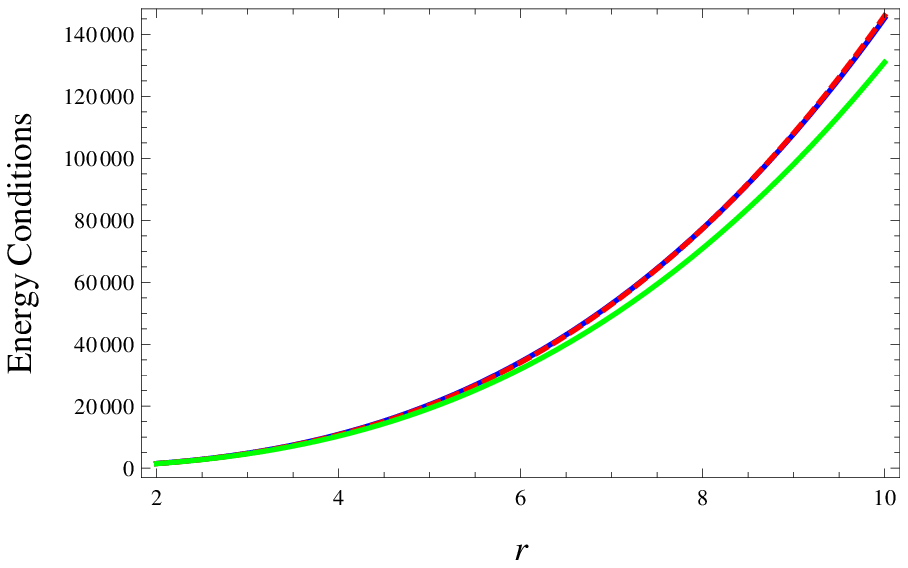, width=0.42\linewidth}\epsfig{file=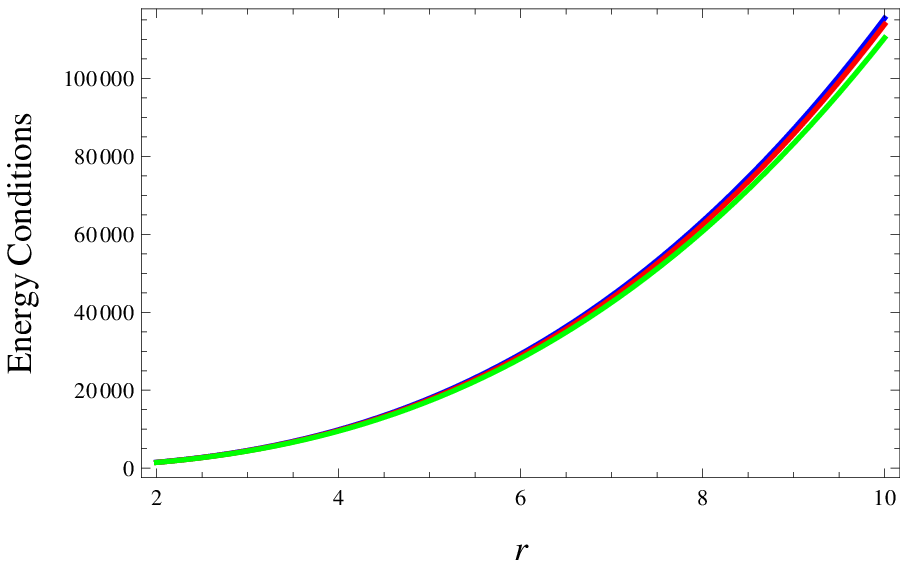,
width=0.42\linewidth}\\\epsfig{file=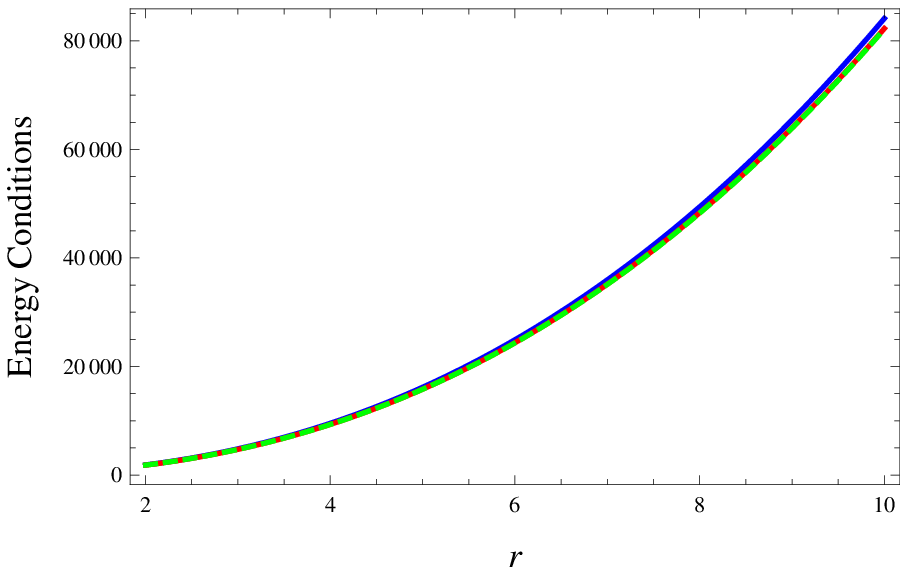,
width=0.42\linewidth}\caption{Plots of $\rho-\mathcal{A}$ (blue),
$\rho+P_{r}-\mathcal{A}$ (red) and $\rho+P_{t}-\mathcal{A}$ (green)
versus $r$ with $\Upsilon=0.05,~\mu=-1,~\nu=1,~a=-0.01$ and
$b=-0.25$. The upper case for $n=1$ (left) and $n=0.5$ (right) while
lower case for $n=-0.5$.}
\end{figure}

\subsection{Isotropic Fluid}

For isotropic matter configuration $P=P_{r}=P_{t}$, Eqs.(\ref{11a})
and (\ref{12a}) lead to
\begin{eqnarray}\nonumber
&&[\Upsilon(1-\Upsilon)+(1+\Upsilon)(2+3\Upsilon)]\left[\frac{2\lambda}
{r^3}\left(1-\frac{\chi}{r}\right)-\frac{\chi}{r^3}+\frac{1}{2}\mu
\mathcal{G}^{a}(1+\nu\mathcal{G}^{b})\right.\\\nonumber&-&\left.
\frac{1}{2}\mu\mathcal{G}^{a}[a+\nu(a+b)\mathcal{G}^{b}]-\frac{4\mu
\lambda}{r^5}(3\chi-2r)\left(1-\frac{\chi}{r}\right)\left\{a(a-1)
+\nu(a+b)\right.\right.\\\nonumber&\times&\left.\left.(a+b-1)
\mathcal{G}^{b}\right\}\mathcal{G}^{a-2}\mathcal{G}'\right]+2
[(1-\Upsilon)^{2}-\Upsilon(1+\Upsilon)]\left[\frac{\lambda}{r^4}
(\lambda-r)\left(1-\frac{\chi}{r}\right)\right.\\\nonumber&+&\left.
\frac{1}{2r^4}(r+\lambda)(\chi-r\chi')+\frac{1}{2}\mu\mathcal{G}^{a}
(1+\nu\mathcal{G}^{b})-\frac{1}{2}\mu\mathcal{G}^{a}[a+\nu(a+b)
\mathcal{G}^{b}]\right.\\\nonumber&+&\left.\frac{2\lambda\mu}{r^5}
\left(1-\frac{\chi}{r}\right)\left\{2(\lambda-2r)\left(1-\frac{\chi}
{r}\right)-3(r\chi'-\chi)\right\}\left[a(a-1)+\nu(a+b)\right.\right.
\\\nonumber&\times&\left.\left.(a+b-1)\mathcal{G}^{b}\right]
\mathcal{G}^{a-2}\mathcal{G}'+\frac{4\lambda}{r^3}\left(1-\frac{\chi}
{r}\right)^{2}\left\{a\mu(a-1)\mathcal{G}^{a-2}\left(\mathcal{G}''
+(a-2)\right.\right.\right.\\\nonumber&\times&\left.\left.\left.
\frac{\mathcal{G}'^{2}}{\mathcal{G}}\right)+\mu\nu(a+b)(a+b-1)
\mathcal{G}^{a+b-2}\left(\mathcal{G}''+(a+b-2)\frac{\mathcal{G}'^{2}}
{\mathcal{G}}\right)\right\}\right]=0,\\\label{1d}
\end{eqnarray}
\begin{figure}\centering
\epsfig{file=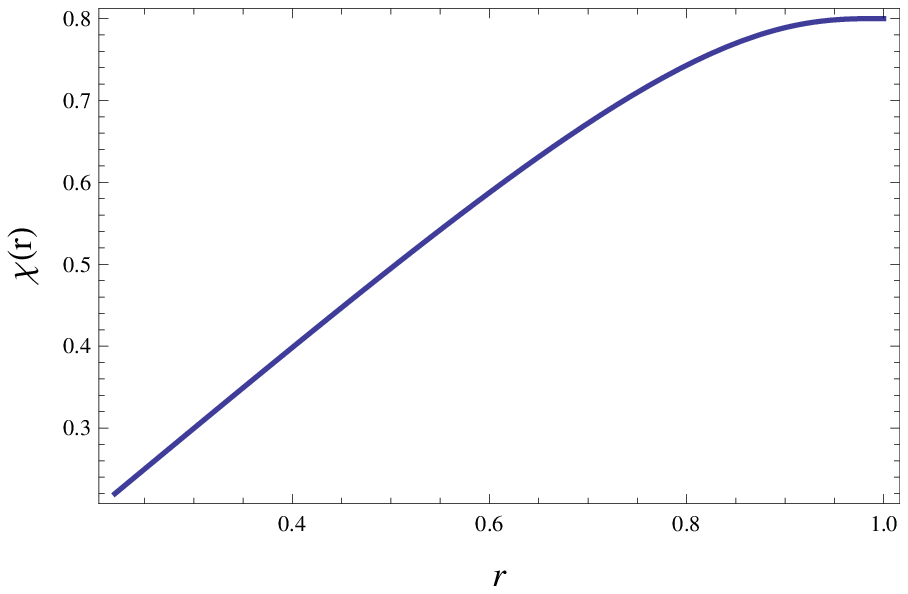, width=0.5\linewidth}\epsfig{file=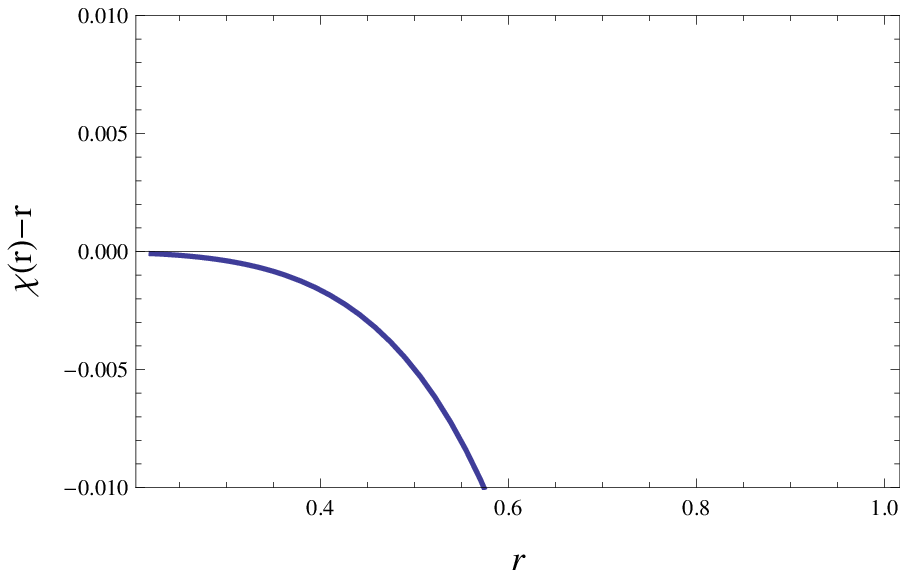,
width=0.5\linewidth}\\\epsfig{file=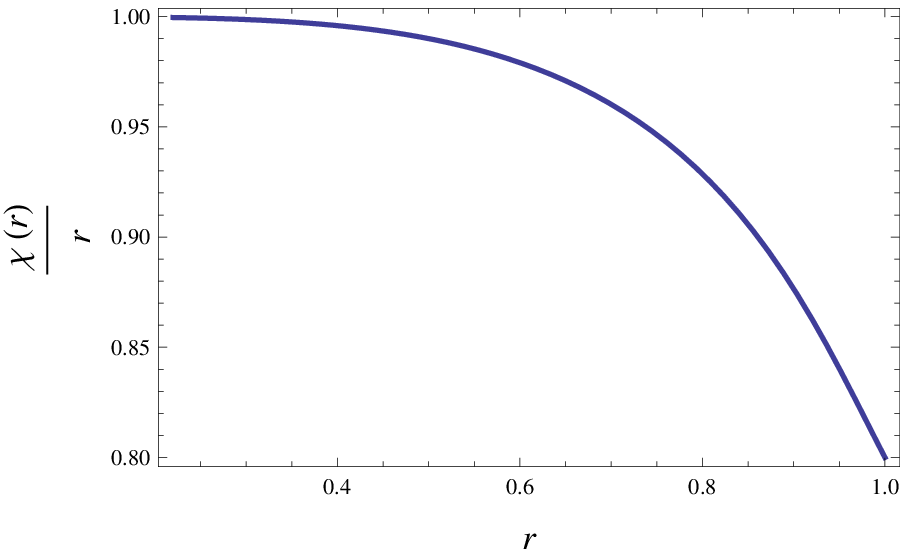, width=0.5\linewidth}
\epsfig{file=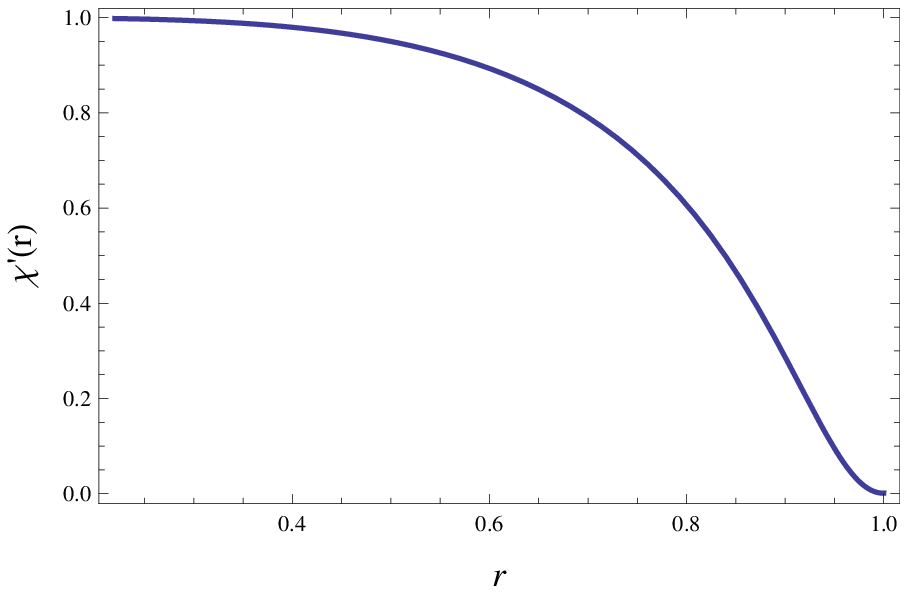, width=0.45\linewidth}\caption{Plots of
$\chi(r),~\chi(r)-r,~\frac{\chi(r)}{r}$ and $\chi'(r)$ versus $r$
for $\mu=1,~\nu=-1,\newline a=0.2,~b=0.25,~\lambda=0.01$ and
$\Upsilon=0.05$.}
\end{figure}
\begin{figure}\centering
\epsfig{file=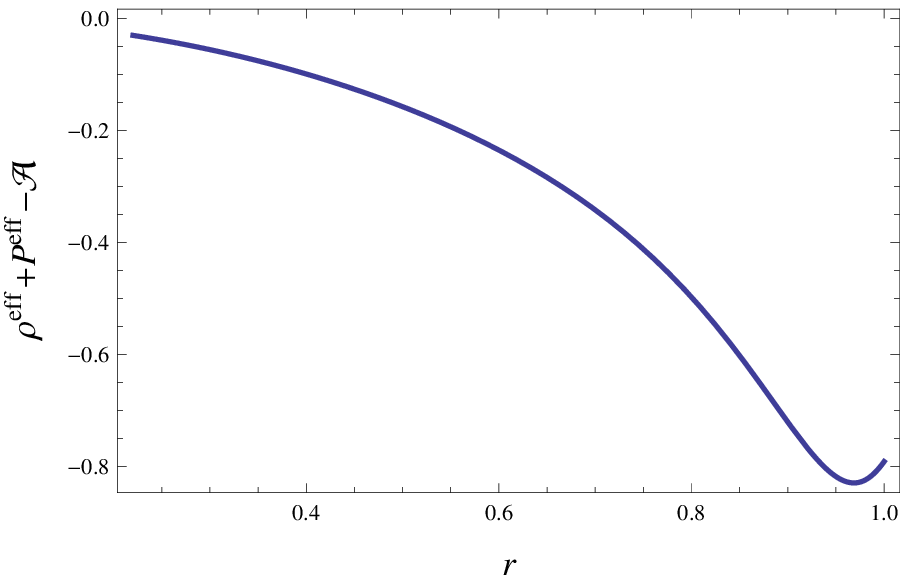, width=0.5\linewidth}\epsfig{file=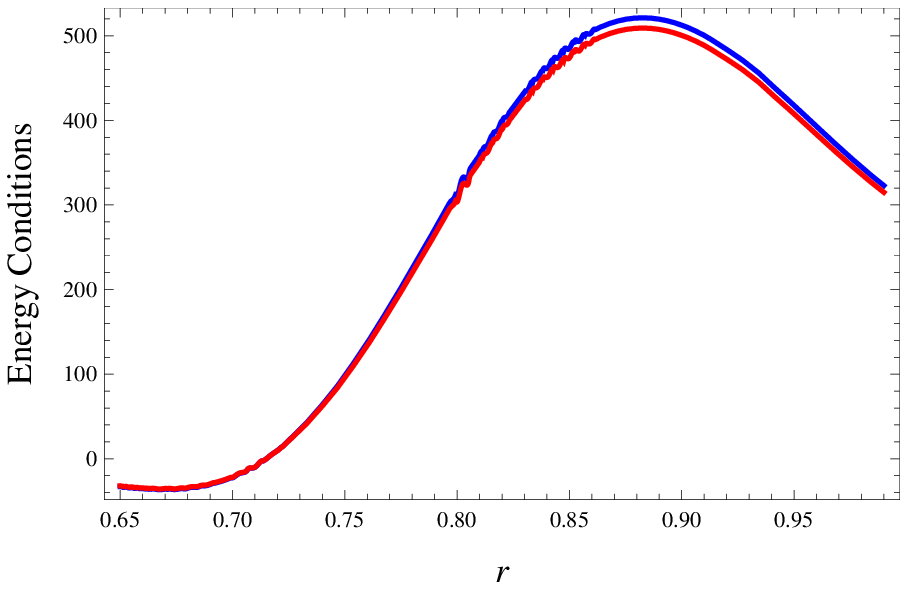,
width=0.5\linewidth}\caption{Plots of energy conditions for same
values.}
\end{figure}
where the value of $\mathcal{G}$ is given in Eq.(\ref{2b}). This
differential equation is highly non-linear in $\chi(r)$ which cannot
be solved analytically. To find its solution, we use the numerical
technique and display the corresponding results in Figure
\textbf{6}. The graphical behavior of $\chi(r)$ is given in the
upper left panel which shows that the positivity of
$\left(1-\frac{\chi}{r}\right)$ is fulfilled throughout the
evolution. The right panel indicates that the WH throat is located
approximately at $r_{\mathrm{th}}=0.2239$ for which $\chi(r)-r$
approaches to zero but fails to cross the radial axis to get the
exact value of $r_{\mathrm{th}}$. The asymptotically flatness
condition is not satisfied as given in the lower left case while the
right plot shows that the condition $\chi'(r_{\mathrm{th}})<1$ is
obeyed. In Figure \textbf{7} (left), the violating behavior of
effective NEC indicates that $f(\mathcal{G},T)$ gravity provides
$T_{\alpha\beta}^{\mathrm{eff}}$ as a required alternative source
while the right panel shows the plots of $\rho-\mathcal{A}$ (blue)
and $\rho+P-\mathcal{A}$ (red). It is observed that both NEC as well
as WEC are positive in the region $0.715\leq r<1$ and hence
realistic traversable tiny WH can be formed for isotropic fluid.

\subsection{Barotropic Fluid}

In this case, we explore the WH geometries using barotropic equation
of state for both radial as well as tangential pressures. For radial
pressure, we take $P_{r}=w\rho$ ($w$ is an equation of state
parameter), Eqs.(\ref{10a}) and (\ref{11a}) give
\begin{eqnarray}\nonumber
&&[2w+(1+5w)\Upsilon]\left[\frac{\chi'}{r^2}-\frac{1}{2}\mu\mathcal{G}^{a}
(1+\nu\mathcal{G}^{b})+\frac{1}{2}\mu\mathcal{G}^{a}[a+\mu(a+b)
\mathcal{G}^{b}]\right.\\\nonumber&+&\left.\frac{2\mu}{r^4}(r\chi'-\chi)
\left(2-\frac{3\chi}{r}\right)\left[a(a-1)+\nu(a+b)(a+b-1)\mathcal{G}^{b}
\right]\mathcal{G}^{a-2}\mathcal{G}'\right.\\\nonumber&+&\left.\frac{4\chi}
{r^3}\left(1-\frac{\chi}{r}\right)\left[a\mu(a-1)\mathcal{G}^{a-2}\left\{
\mathcal{G}''+(a-2)\frac{\mathcal{G}'^{2}}{\mathcal{G}}\right\}+\mu\nu(a+b)
(a+b-1)\right.\right.\\\nonumber&\times&\left.\left.\mathcal{G}^{a+b-2}\left\{
\mathcal{G}''+(a+b-2)\frac{\mathcal{G}'^{2}}{\mathcal{G}}\right\}\right]
\right]-[2-(w-3)\Upsilon]\left[\frac{2\lambda}{r^3}\left(1-\frac{\chi}{r}
\right)-\frac{\chi}{r^3}\right.\\\nonumber&+&\left.\frac{1}{2}\mu
\mathcal{G}^{a}(1+\nu\mathcal{G}^{b})-\frac{1}{2}\mu\mathcal{G}^{a}
[a+\nu(a+b)\mathcal{G}^{b}]-\frac{4\mu\lambda}{r^5}(3\chi-2r)\left(1
-\frac{\chi}{r}\right)\right.\\\nonumber&\times&\left.\left\{a(a-1)+\nu
(a+b)(a+b-1)\mathcal{G}^{b}\right\}\mathcal{G}^{a-2}\mathcal{G}'\right]
+2\Upsilon(1+w)\left[\frac{\lambda}{r^4}(\lambda-r)\right.\\\nonumber
&\times&\left.\left(1-\frac{\chi}{r}\right)+\frac{1}{2r^4}(r+\lambda)
(\chi-r\chi')+\frac{1}{2}\mu\mathcal{G}^{a}(1+\nu\mathcal{G}^{b})
-\frac{1}{2}\mu\mathcal{G}^{a}[a+\nu(a+b)\right.\\\nonumber&\times&\left.
\mathcal{G}^{b}]+\frac{2\lambda\mu}{r^5}\left(1-\frac{\chi}{r}\right)
\left\{2(\lambda-2r)\left(1-\frac{\chi}{r}\right)-3(r\chi'-\chi)\right\}
\left[a(a-1)\right.\right.\\\nonumber&+&\left.\left.\nu(a+b)(a+b-1)
\mathcal{G}^{b}\right]\mathcal{G}^{a-2}\mathcal{G}'+\frac{4\lambda}{r^3}
\left(1-\frac{\chi}{r}\right)^{2}\left\{a\mu(a-1)\mathcal{G}^{a-2}\right.
\right.\\\nonumber&\times&\left.\left.\left(\mathcal{G}''+(a-2)
\frac{\mathcal{G}'^{2}}{\mathcal{G}}\right)+\mu\nu(a+b)(a+b-1)
\mathcal{G}^{a+b-2}\right.\right.\\\label{1f}&\times&\left.\left.
\left(\mathcal{G}''+(a+b-2)\frac{\mathcal{G}'^{2}}{\mathcal{G}}\right)
\right\}\right]=0,
\end{eqnarray}
which we solve numerically for $\chi(r)$ and present the results in
Figure \textbf{8}. The positively increasing evolution of $\chi(r)$
is shown in the upper left case while the right plot gives the value
of WH throat radius. In this case, the throat is located at
$r_{\mathrm{th}}=0.1572$ such that
$\chi(r_{\mathrm{th}})=r_{\mathrm{th}}$. The lower panel (left)
indicates non-asymptotically flat behavior of $\chi(r)$ whereas the
constraint $\chi'(r)<1$ is satisfied at $r=r_{\mathrm{th}}$ (right).
The left plot of Figure \textbf{9} indicates negativity of NEC for
$T_{\alpha\beta}^{\mathrm{eff}}$ while the right panel shows that
NEC as well as WEC for ordinary matter distribution described by
$\rho-\mathcal{A}\geq0$ (blue), $\rho+P_{r}-\mathcal{A}\geq0$ (red)
and $\rho+P_{t}-\mathcal{A}\geq0$ (green) are satisfied in the
region $0.66\leq r\leq 0.90$. Thus, the WH geometries are supported
by ordinary matter in this case.
\begin{figure}\centering
\epsfig{file=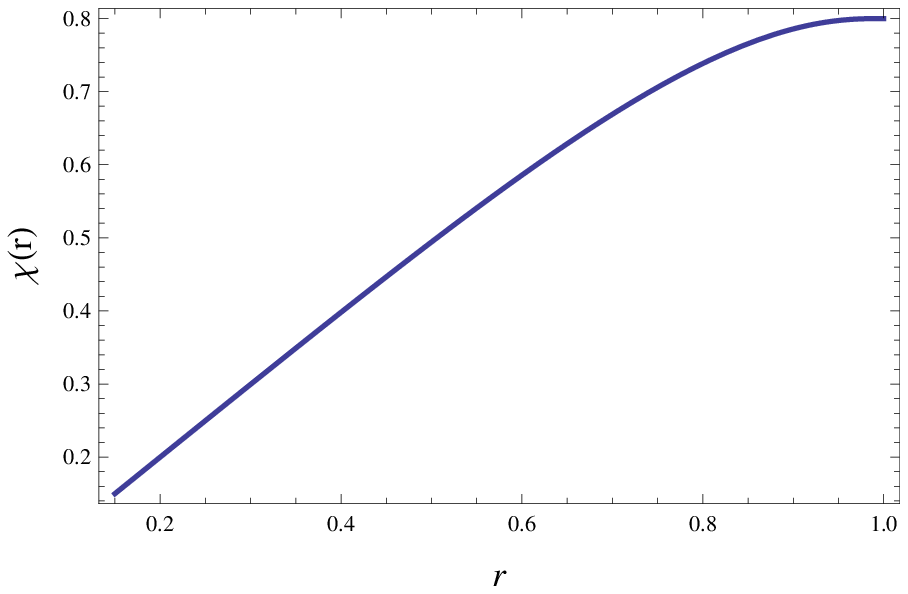, width=0.5\linewidth}\epsfig{file=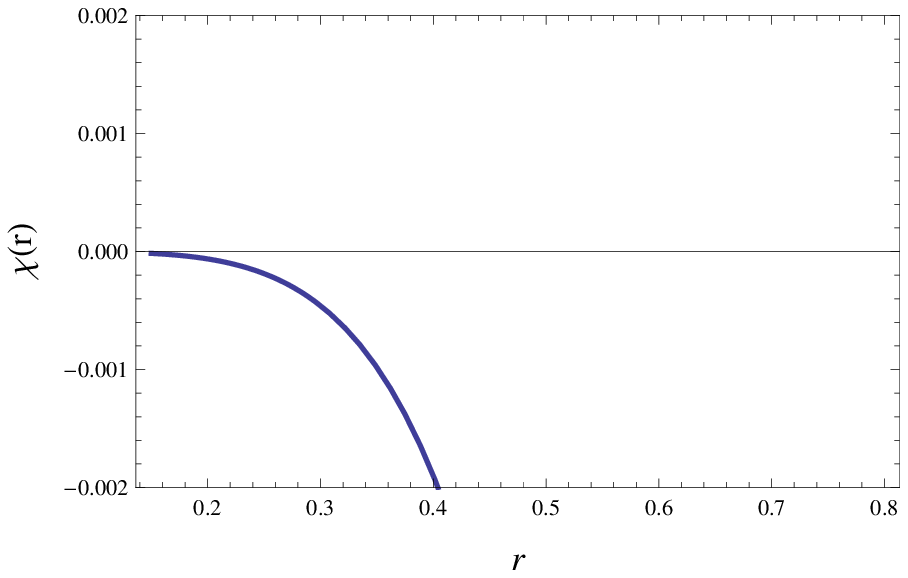,
width=0.5\linewidth}\\\epsfig{file=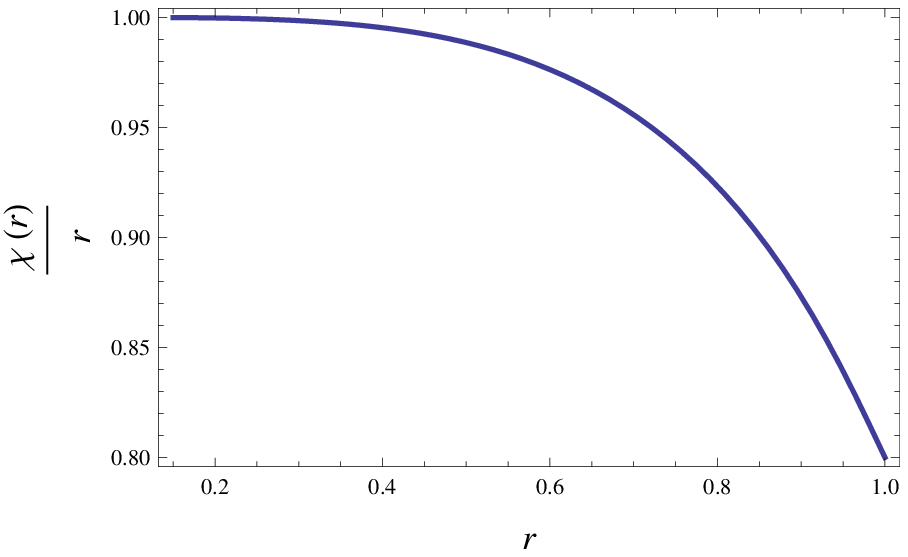, width=0.5\linewidth}
\epsfig{file=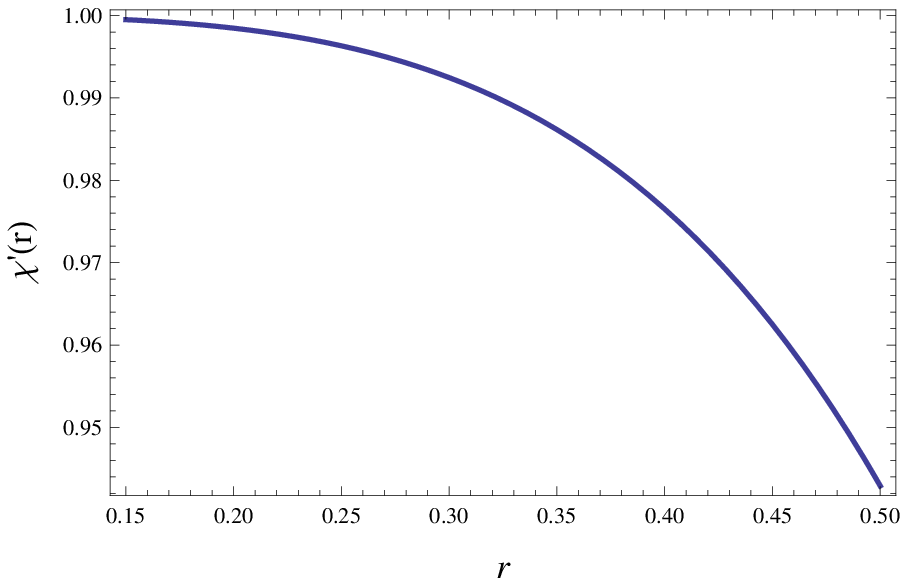, width=0.45\linewidth}\caption{Plots of
$\chi(r),~\chi(r)-r,~\frac{\chi(r)}{r}$ and $\chi'(r)$ versus $r$
for $w=-0.1$.}
\end{figure}
\begin{figure}\centering
\epsfig{file=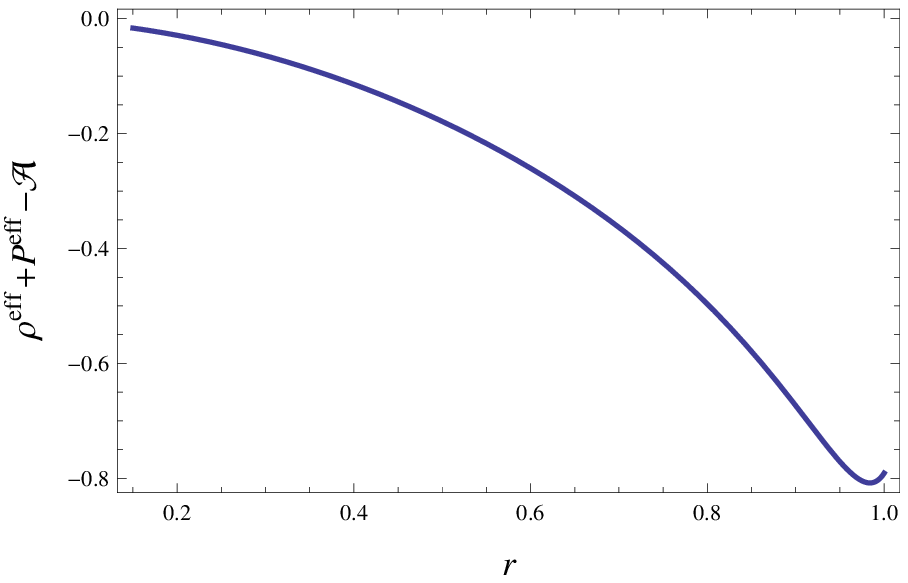, width=0.5\linewidth}\epsfig{file=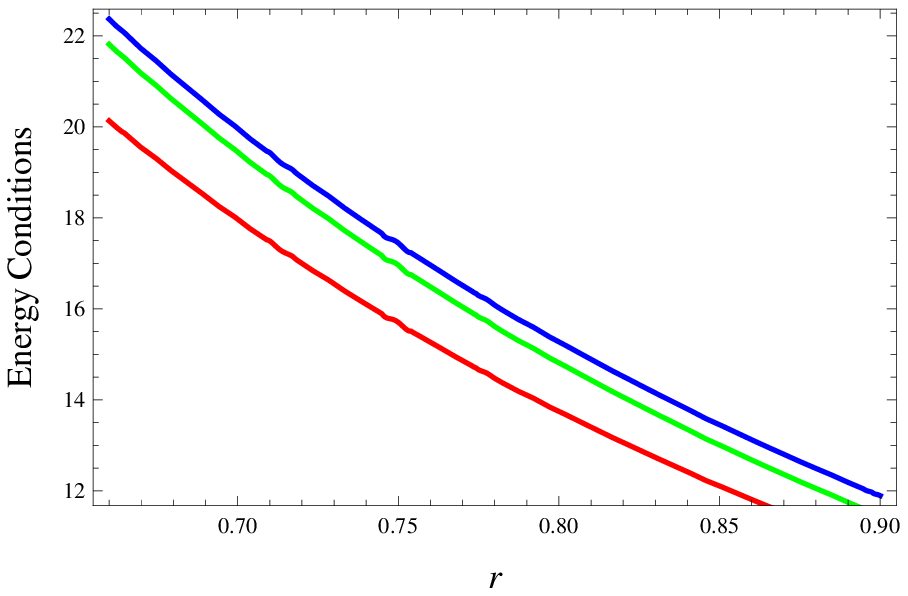,
width=0.5\linewidth}\caption{Plots of energy conditions for same
values.}
\end{figure}

In case of tangential pressure, we consider the barotropic equation
of state of the form $P_{t}=w\rho$. To analyze the possible
existence of WH solutions, Eqs.(\ref{10a}) and (\ref{12a}) give
third order differential equation in $\chi(r)$ as
\begin{eqnarray}\nonumber
&&(1+\Upsilon)[2w+(1+5w)\Upsilon]\left[\frac{\chi'}{r^2}-\frac{1}{2}
\mu\mathcal{G}^{a}(1+\nu\mathcal{G}^{b})+\frac{1}{2}\mu\mathcal{G}^{a}
[a+\mu(a+b)\right.\\\nonumber&\times&\left.\mathcal{G}^{b}]
+\frac{2\mu}{r^4}(r\chi'-\chi)\left(2-\frac{3\chi}{r}\right)\left[a
(a-1)+\nu(a+b)(a+b-1)\mathcal{G}^{b}\right]\right.\\\nonumber&\times&
\left.\mathcal{G}^{a-2}\mathcal{G}'+\frac{4\chi}{r^3}\left(1
-\frac{\chi}{r}\right)\left[a\mu(a-1)\mathcal{G}^{a-2}\left\{
\mathcal{G}''+(a-2)\frac{\mathcal{G}'^{2}}{\mathcal{G}}\right\}
\right.\right.\\\nonumber&+&\left.\left.\mu\nu(a+b)(a+b-1)
\mathcal{G}^{a+b-2}\left\{\mathcal{G}''+(a+b-2)\frac{
\mathcal{G}'^{2}}{\mathcal{G}}\right\}\right]\right]+\Upsilon
[(1+w)\\\nonumber&-&(1-w)\Upsilon]\left[\frac{2\lambda}{r^3}
\left(1-\frac{\chi}{r}\right)-\frac{\chi}{r^3}+\frac{1}{2}\mu
\mathcal{G}^{a}(1+\nu\mathcal{G}^{b})-\frac{1}{2}\mu\mathcal{G}^{a}
[a+\nu(a+b)\right.\\\nonumber&\times&\left.\mathcal{G}^{b}]
-\frac{4\mu\lambda}{r^5}(3\chi-2r)\left(1-\frac{\chi}{r}\right)
\left\{a(a-1)+\nu(a+b)(a+b-1)\mathcal{G}^{b}\right\}\right.
\\\nonumber&\times&\left.\mathcal{G}^{a-2}\mathcal{G}'\right]
+2[w\Upsilon(1+\Upsilon)+(1-\Upsilon)^{2}]\left[\frac{\lambda}
{r^4}(\lambda-r)\left(1-\frac{\chi}{r}\right)+\frac{1}{2r^4}
(r+\lambda)\right.\\\nonumber&\times&\left.(\chi-r\chi')+\frac{1}
{2}\mu\mathcal{G}^{a}(1+\nu\mathcal{G}^{b})-\frac{1}{2}\mu
\mathcal{G}^{a}[a+\nu(a+b)\mathcal{G}^{b}]+\frac{2\lambda\mu}
{r^5}\left(1-\frac{\chi}{r}\right)\right.\\\nonumber&\times&
\left.\left\{2(\lambda-2r)\left(1-\frac{\chi}{r}\right)-3(r
\chi'-\chi)\right\}\left[a(a-1)+\nu(a+b)(a+b-1)\mathcal{G}^{b}
\right]\right.\\\nonumber&\times&\left.\mathcal{G}^{a-2}
\mathcal{G}'+\frac{4\lambda}{r^3}\left(1-\frac{\chi}{r}
\right)^{2}\left\{a\mu(a-1)\mathcal{G}^{a-2}\left(\mathcal{G}''
+(a-2)\frac{\mathcal{G}'^{2}}{\mathcal{G}}\right)\right.\right.
\\\label{1g}&+&\left.\left.\mu\nu(a+b)(a+b-1)\mathcal{G}^{a+b-2}
\left(\mathcal{G}''+(a+b-2)\frac{\mathcal{G}'^{2}}{\mathcal{G}}
\right)\right\}\right]=0.
\end{eqnarray}
We solve this equation numerically and corresponding results are
displayed in Figure \textbf{10}. Similar behavior of
$\chi(r),~\frac{\chi(r)}{r}$ and $\chi'(r)$ are obtained as in the
previous cases. The throat is located at $r_{\mathrm{th}}=0.1585$
where the curve $\chi(r)-r$ approaches to zero. The left panel of
Figure \textbf{11} shows the violation of NEC in $f(\mathcal{G},T)$
gravity throughout the evolution. The graphs of
$\rho-\mathcal{A}\geq0$ (blue), $\rho+P_{r}-\mathcal{A}\geq0$ (red)
and $\rho+P_{t}-\mathcal{A}\geq0$ (green) exhibit positive values in
the interval $0.66\leq r\leq 0.90$ as shown in Figure \textbf{11}
(right). Thus, a micro WH can be formed in this physically
acceptable region.
\begin{figure}\centering
\epsfig{file=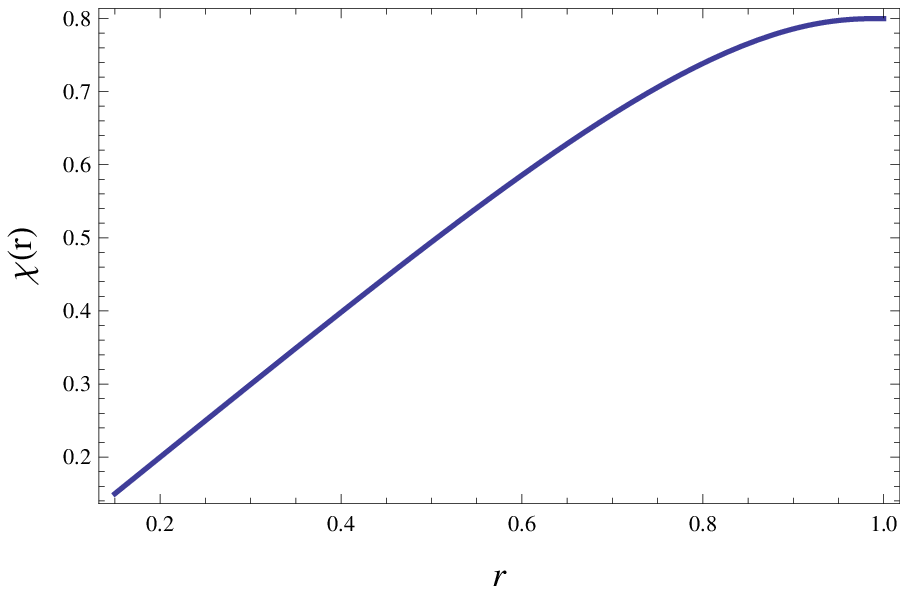, width=0.5\linewidth}\epsfig{file=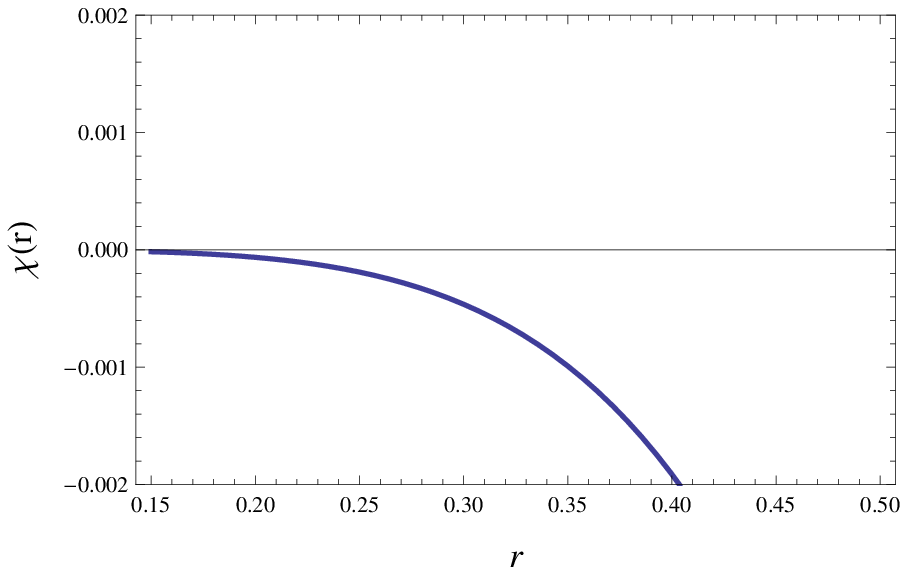,
width=0.5\linewidth}\\\epsfig{file=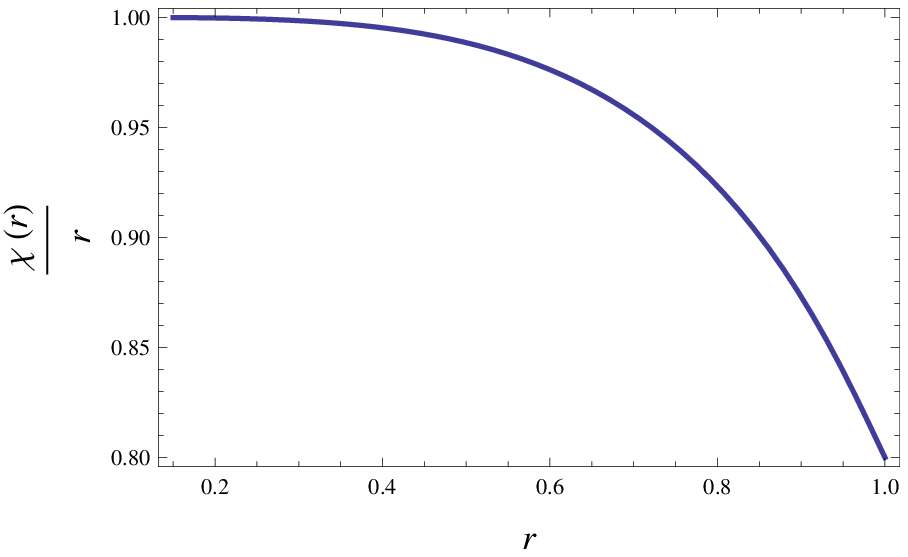, width=0.5\linewidth}
\epsfig{file=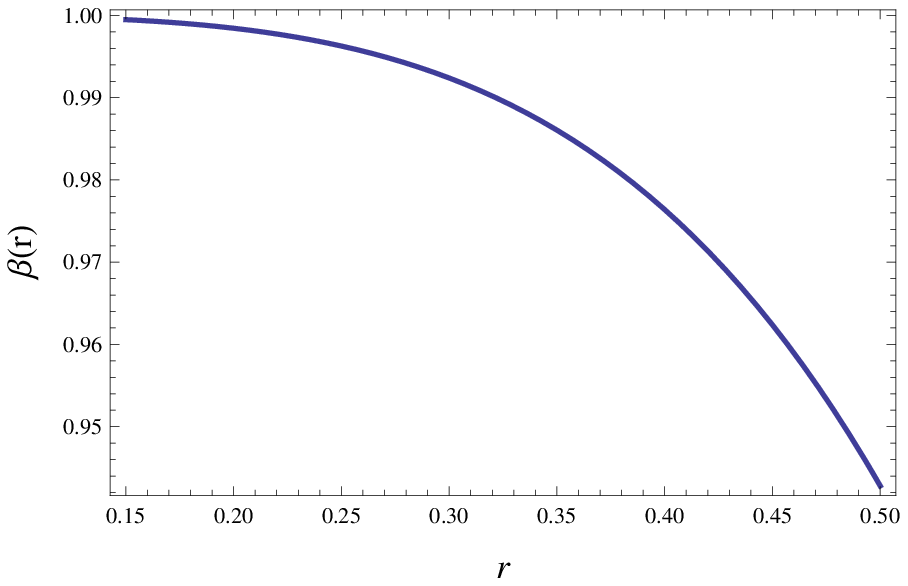, width=0.45\linewidth}\caption{Plots of
$\chi(r),~\chi(r)-r,~\frac{\chi(r)}{r}$ and $\chi'(r)$ versus $r$
for $w=-0.04$.}
\end{figure}
\begin{figure}\centering
\epsfig{file=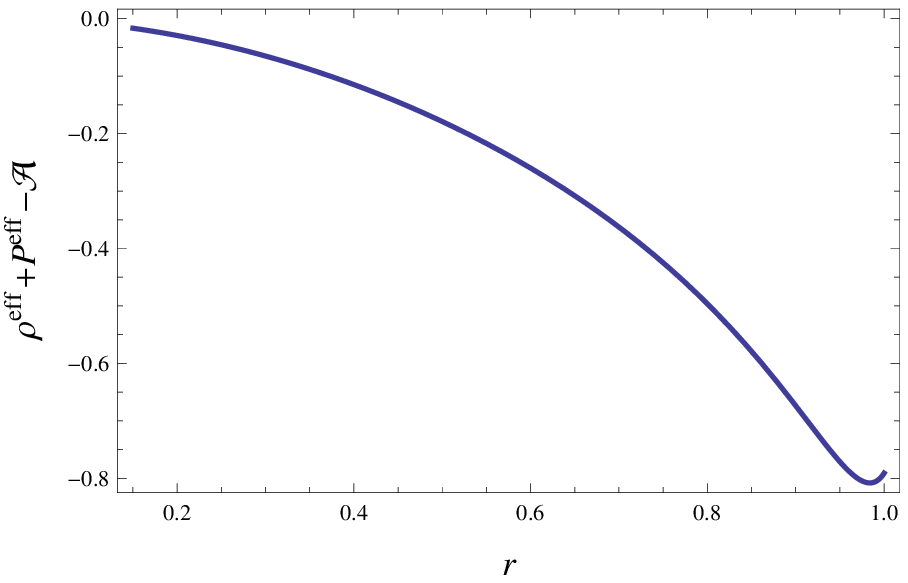, width=0.5\linewidth}\epsfig{file=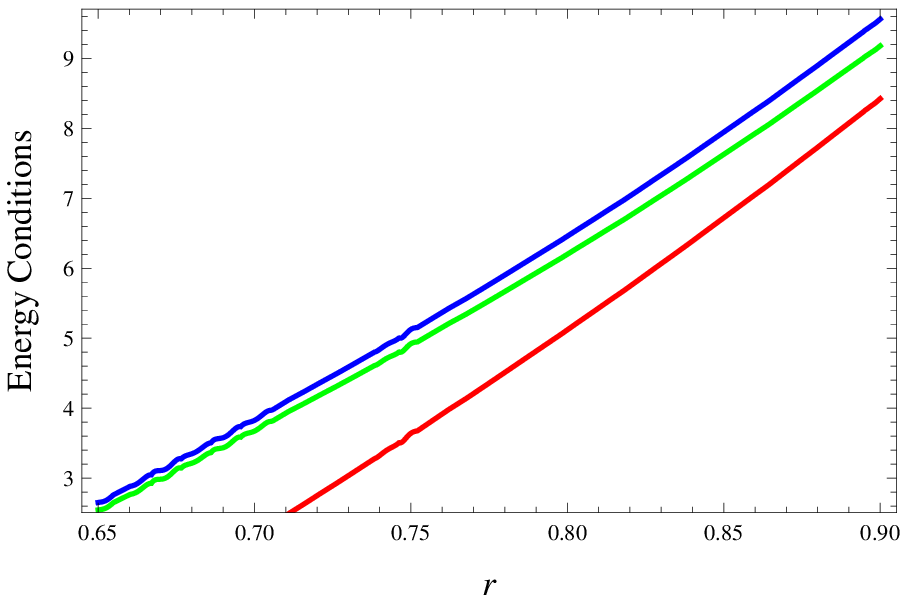,
width=0.5\linewidth}\caption{Plots of energy conditions for same
values.}
\end{figure}
\newpage
\section{Final Remarks}

In this paper, we have analyzed static spherically symmetric
traversable WH solutions in $f(\mathcal{G},T)$ gravity using
anisotropic, isotropic as well as barotropic matter distributions.
For this purpose, we have considered a particular model
$f(\mathcal{G},T)=F(\mathcal{G})+\mathcal{F}(T)$ with specific form
of $\psi(r)$ satisfying the no-horizon condition. In
$f(\mathcal{G},T)$ gravity, test particles follow non-geodesic
trajectories due to the presence of extra force which appears as a
consequence of non-zero divergence of the energy-momentum tensor
\cite{4}. For the non-geodesic congruences, the divergence of
four-acceleration appears in the evolution of expansion scalar
\cite{18}. Consequently, the four fundamental energy conditions for
non-geodesic congruences are also affected due to the presence of
this auxiliary term. We have formulated these non-geodesic energy
constraints to explore the existence of realistic WH solutions in
this gravity.

For anisotropic fluid, we have considered viable form of $\chi(r)$
which meets all necessary conditions of traversable WH geometry and
analyzed the behavior of energy conditions. This analysis is carried
out for four possible choices of $F(\mathcal{G})$ model parameters
such that they avoid all four types of finite-time future
singularities. For isotropic and barotropic (satisfied by radial as
well as tangential pressures) matter distributions, we have solved
the corresponding differential equations numerically to examine the
behavior of $\chi(r)$. The non-asymptotically flat shape functions
satisfying the basic requirements for WH geometry are obtained in
each case. The violation of NEC defined in $f(\mathcal{G},T)$
gravity is observed throughout the evolution for all three types of
matter distribution. This violation confirms the traversability of
WH solutions while the positivity of NEC as well as WEC for ordinary
matter contents assure the existence of physical acceptable WH
geometries in certain regions.

In $f(\mathcal{G})$ gravity, we have found that realistic WH
solutions exist only for barotropic fluid satisfied by radial
pressure \cite{14}. This difference may be due to the
curvature-matter coupling present in $f(\mathcal{G},T)$ gravity. We
can conclude that this curvature-matter coupled theory provides an
alternative source for the existence of realistic WH solutions.

\end{document}